\documentclass[showpacs,
preprint,preprintnumbers,nofootinbib,
groupedaddress,superscriptaddress,amsmath,amssymb]{revtex4}
\usepackage{color}
\usepackage{graphicx}

\newcommand{\lsim}{\mbox{ \raisebox{-1.0ex}{$\stackrel{\textstyle <}
{\textstyle \sim}$ }}}

\allowdisplaybreaks
\def\lq{\em{L}\hspace{-0.6ex}\em{Q}}

\begin{document}
\title{Higgs boson pair production in new physics models\\ at hadron, lepton, and photon colliders}
\preprint{OCHA-PP-305, KEK-TH-1396, UT-HET 041, IC/2010/076}
\author{Eri~Asakawa}
\email{eri@post.kek.jp}
\affiliation{Department of Physics, Ochanomizu University, Tokyo 112-8610, Japan}
\author{Daisuke~Harada}
\email{dharada@post.kek.jp}
\affiliation{KEK Theory Center, Institute of Particle and Nuclear Studies,
KEK, 1-1 Oho, Tsukuba, Ibaraki 305-0801, Japan}
\affiliation{Department of Particle and Nuclear Physics, The Graduate
University for Advanced Studies (Sokendai), 1-1 Oho, Tsukuba, Ibaraki 305-0801, Japan}
\author{Shinya~Kanemura}
\email{kanemu@sci.u-toyama.ac.jp}
\affiliation{Department of Physics, The University of Toyama, 3190 Gofuku, Toyama 930-8555, Japan}
\author{Yasuhiro~Okada}
\email{yasuhiro.okada@kek.jp}
\affiliation{KEK Theory Center, Institute of Particle and Nuclear Studies,
KEK, 1-1 Oho, Tsukuba, Ibaraki 305-0801, Japan}
\affiliation{Department of Particle and Nuclear Physics, The Graduate
University for Advanced Studies (Sokendai), 1-1 Oho, Tsukuba, Ibaraki 305-0801, Japan}
\author{Koji Tsumura}
\email{ktsumura@ictp.it}
\affiliation{The Abdus Salam ICTP of
UNESCO and IAEA, Strada Costiera 11, 34151 Trieste, Italy}
\begin{abstract}
We study Higgs boson pair production processes
at future hadron and lepton colliders including the photon collision
option in several new physics models; i.e., the two-Higgs-doublet model,
the scalar leptoquark model, the sequential fourth generation fermion model
and the vectorlike quark model.
Cross sections for these processes can deviate significantly from
the standard model predictions due to the one-loop correction to
the triple Higgs boson coupling constant. For the one-loop
induced processes such as $gg \to hh$ and $\gamma\gamma\to hh$, where $h$ is
the (lightest) Higgs boson and $g$ and $\gamma$ respectively represent a gluon
and a photon, the cross sections can also be affected by new physics particles
via additional one-loop diagrams.
In the two-Higgs-doublet model and scalar leptoquark models, cross sections of
$e^+e^-\to hhZ$ and $\gamma\gamma\to hh$ can be enhanced due to the nondecoupling
effect in the one-loop corrections to the triple Higgs boson coupling constant.
In the sequential fourth generation fermion model, the cross section for $gg\to hh$
becomes very large because of the loop effect of the fermions.
In the vectorlike quark model, effects are small because the theory has decoupling
property.
Measurements of the Higgs boson pair production processes can be useful to explore
new physics through the determination of the Higgs potential.
\end{abstract}
\maketitle

\section{Introduction}

The standard model (SM) for particle physics has experienced a great success in
describing the experimental data of high energy physics below the energy range of
a hundred GeV, but its portion for electroweak symmetry breaking, the Higgs sector,
remains unknown.  Experimental confirmation of the Higgs boson is one of the most
important issues in the high energy physics.
The direct search results at the LEP experiment have constrained the mass ($m_h$)
of the Higgs boson as $m_h\gtrsim 114.4$~GeV~\cite{Ref:LEP} in the SM with one
Higgs doublet, and the global analysis of precision measurements for electroweak
observables has indicated that $m_h$ is smaller than $157$ GeV at the $95\%$
confidence level~\cite{Ref:LEP,Ref:LEPEW}. In addition, the combined data from
the CDF and D0 collaborations at the Fermilab Tevatron have excluded the region
of $162$ GeV $\lesssim m_h \lesssim 166$ GeV~\cite{Ref:mh165GeV}
\footnote{
Recently, the bound on the Higgs boson mass from Tevatron experiments
has been updated\cite{ICHEPslide}.}.
The CERN Large Hadron Collider (LHC) has already started its operation, and it
will soon be ready for hunting the Higgs boson. We expect that the Higgs boson
will be discovered in coming several years.

Once the Higgs boson is found at the Tevatron or the LHC, its property such as
the mass, the decay width, production cross sections and the decay branching ratios
will be thoroughly measured as accurately as possible in order to confirm whether
it is really the particle responsible for spontaneous breaking of the electroweak
symmetry. The Higgs mechanism will be tested by determining the coupling constants
of the Higgs boson to the weak gauge bosons. The measurement of the Yukawa coupling
constants will clarify the mass generation mechanism of quarks and charged leptons.
However, in order to understand the physics behind the electroweak symmetry breaking,
the Higgs potential must be reconstructed by measuring the triple Higgs boson coupling
constant (the $hhh$ coupling constant).

On the other hand, from the theoretical view point, it would be expected that the SM
is replaced by a more fundamental theory at the TeV scale. One way to see the new
dynamics is to measure effective vertices of the SM fields and to compare them to
the theoretical calculation of radiative corrections.
The effect can be significant in the electroweak theory especially when the mass of
a new particle comes mainly from the vacuum expectation value (VEV) of the Higgs field
like chiral fermions. In such a case the decoupling theorem~\cite{decoupling} does not
necessarily hold, so that the new physics effects do not decouple and are significant.
It is well known that the systematic study of nondecoupling parameters in radiative
corrections to the gauge boson two point functions has played an important role to
constrain new physics models by using the precision data of electroweak observables
at the LEP and the Stanford Linear Collider (SLC)\cite{stu}.

Such nondecoupling effects of new physics particles can also be very significant
in the radiative corrections to the $hhh$ coupling constant~\cite{Ref:KOSY}. Quartic
powerlike contributions of the mass of a new particle can appear in the one-loop
correction to the $hhh$ coupling constant, which can give a large deviation from the
SM prediction. For example, in the two-Higgs-doublet model (THDM), the $hhh$ coupling
constant of the lightest (SM-like) Higgs boson can be deviated by ${\mathcal O}(100)\%$
due to nondecoupling effects of extra scalar bosons in radiative corrections without
contradiction with perturbative unitarity~\cite{Ref:KOSY}. It is known that such a large
deviation in the $hhh$ coupling constant from the SM value can be a common feature of
the Higgs sector with the strong first order electroweak phase
transition~\cite{1OPT,1OPT11,1OPT12,1OPT2,1OPT3}, which is required for a successful
scenario of electroweak baryogenesis~\cite{Ref:EWBG}. Therefore, the measurement of
the $hhh$ coupling constant at collider experiments can be an important probe into
such a cosmological scenario.
The one-loop contributions to the $hhh$ coupling constant can also be very large in
the model with sequential fourth generation fermions~\cite{Ref:Kribs4G} and a class
of extended supersymmetric SMs~\cite{Ref:KSY}.
%

At the LHC, the measurement of the $hhh$ coupling constant would be challenging.
In the SM, the cross section of double Higgs boson production from gluon fusion,
$gg\to hh$~\cite{Ref:gghh-qqhh,Ref:gghh,Ref:gghh-Vhh-gghh}, can be ${\mathcal O}(10)$ fb
for $m_h=120$--$160$ GeV with the collision energy of $\sqrt{s}=14$ TeV, while those
of double-Higgs-strahlung $q\bar q\to V^*\to hhV$\cite{Ref:Vhh,Ref:gghh-Vhh-gghh} and
vector boson fusion
$q\bar q\to V^*V^*q\bar q\to hhq\bar q$~\cite{Ref:gghh-qqhh,Ref:qqhh,Ref:gghh-Vhh-gghh}
are much smaller. The double Higgs boson production mechanism from gluon fusion has
been studied in Ref.~\cite{Ref:hhh-sensitivity} with the $h\to WW^{(*)}$ decay mode.
They conclude that the luminosity of $3000$ fb$^{-1}$ is required to measure
the $hhh$ coupling constant at the $20$--$30$\% level~\cite{Ref:hhh-sensitivity}.
For a light Higgs boson ($m_h\lesssim 130$ GeV), the main decay mode is $h\to b\bar b$
which cannot be useful due to huge QCD backgrounds, so that the $hhh$ coupling constant
cannot be accurately measured at the LHC.
%

At the International Linear Collider (ILC), the accuracy for measuring the $hhh$ coupling
constant would be better than that at the LHC depending on the mass of the Higgs boson.
A Higgs boson pair can be produced in the double-Higgs-strahlung process
$e^+e^-\to hhZ$~\cite{Ref:hhZ} and the $W$ fusion mechanism
$e^+e^-\to hh \nu\bar\nu$~\cite{Ref:hhnunu}. At the first stage of the ILC where $e^+e^-$
energy is $500$ GeV the $hhh$ coupling can be measured via the double-Higgs-strahlung
process for $m_h\lesssim 140$ GeV~\cite{Ref:hhZ2,Ref:Yasui,Ref:hhh-sensitivity-ilc,Ref:Baur}.
The evaluation of the statistical sensitivity for the $hhh$ coupling constant is about
$20\%$ accuracy~\cite{Ref:hhZ2,Ref:Yasui}. Detailed simulation studies for this process
are ongoing, which shows that the sensitivity may be lower~\cite{Ref:hhZsim}.
The photon linear collider (PLC) option may also be useful to explore the $hhh$ coupling
constant for $120\lesssim m_h \lesssim 200$ GeV~\cite{Ref:hhhPLC,Ref:harada}.
A simulation study is also in progress~\cite{Ref:hhhPLCsim}.
At the second stage of the ILC ($\sqrt{s}=1$ TeV) or the Compact Linear Collider (CLIC)
where the collision energy would be at a multi-TeV scale, the double Higgs boson pair
production from $W$ boson fusion becomes important because the cross section is larger
due to the $t$-channel enhancement~\cite{Ref:Yasui,Ref:Baur}. The statistical sensitivity
for the $hhh$ coupling constant is less than $10\%$~\cite{Ref:Yasui}.

In this paper, we study how the $hhh$ coupling constant affects cross sections for
the double Higgs boson production processes $gg\to hh$, $e^+e^-\to hhZ$,
$e^+e^-\to hh\nu\bar\nu$ and $\gamma\gamma\to hh$ in various new physics models such as
the THDM, models with scalar leptoquarks, the model with the chiral fourth generation
fermions and the model with vectorlike quarks.
Cross sections for these Higgs boson pair production processes are evaluated, and
can deviate significantly from the SM predictions due to the deviation in the one-loop
corrected $hhh$ coupling constant. In these processes, the effect of the deviation
in the $hhh$ coupling constant mainly appears in the interference of the diagram with
the $hhh$ coupling constant and the other diagrams. Thus, the sign of the deviation
can be important. Also, in the one-loop induced processes such as $gg\to hh$ and
$\gamma\gamma\to hh$, cross sections can depend on new physics particles in additional
one-loop diagrams.
In the THDM and scalar leptoquark models, cross sections for $e^+e^-\to hhZ$ and
$\gamma\gamma\to hh$ can be enhanced due to the nondecoupling effect on the $hhh$
coupling constant through the extra scalar loops. In the chiral fourth generation model,
cross sections of double Higgs boson production processes can become significantly large,
because new particles mediate in the leading order loop diagram as well as the nondecoupling
effect on the $hhh$ coupling constant. In models with vectorlike quarks, the effect on
the cross sections are small because of the decoupling nature of the theory.
By measuring these double Higgs boson production processes at different future collider
experiments, we would be able to test properties of new physics particles in the loop,
which helps identify the new physics model.

In Sec.~II, effects of the $hhh$ coupling constant in Higgs boson pair production processes
$gg\to hh$ at LHC, $e^+e^-\to hhZ$ and $e^+e^-\to hh\nu\bar\nu$ at ILC and CLIC, and
$\gamma\gamma\to hh$ at their photon collider options are discussed. Model dependent analyses
for these processes are given in Sec.~III for the THDM, the scalar leptoquark models,
the chiral fourth generation model, and the vectorlike quarks. In Sec.~IV, summary and
discussions are given.

\section{The Higgs boson pair production processes at colliders}
In this section, we discuss Higgs boson pair production processes
$gg\to hh$~\cite{Ref:gghh-qqhh,Ref:gghh,Ref:gghh-Vhh-gghh}, $e^+e^-\to hhZ$~\cite{Ref:hhZ},
$e^+e^-\to hh\nu\bar\nu$~\cite{Ref:hhnunu} and $\gamma\gamma\to hh$~\cite{Ref:hhhPLC}
in various new physics models. These processes contain the $hhh$ coupling constant so that
they can be used to determine the $hhh$ coupling constant at future collider experiments.
The effective $ggh$ and $\gamma\gamma h$ vertices would be precisely measured in the single
Higgs boson production processes as $gg\to h$ at hadron colliders~\cite{Ref:ggh} and
$\gamma\gamma\to h$ resonance production at the PLC~\cite{Ref:AAh}, which will be used
to extract the $hhh$ coupling constant from the one-loop induced processes such as $gg\to hh$
and $\gamma\gamma\to hh$.
In this section, before going to the discussion on the calculation for the cross sections
in each model, we first consider the results in the SM with a constant shift of the $hhh$
coupling constant by a factor of $(1+\Delta \kappa)$;
\begin{align}
\lambda_{hhh} = \lambda_{hhh}^{\rm SM} (1+\Delta \kappa),
\end{align}
where $\lambda_{hhh}^{\rm SM}=-3m_h^2/v$ at the tree level
\footnote{At the one-loop order, the effective $hhh$ vertex function have been
evaluated as~\cite{Ref:KOSY}
\begin{align}
\Gamma_{hhh}^\text{SM}({\hat s},m_h^2,m_h^2)
&\simeq -\frac{3m_h^2}{v}\left\{1-\frac{N_cm_t^4}{3\pi^2v^2m_h^2}\left[1
+ {\cal O}\left(\frac{m_h^2}{m_t^2}, \frac{\hat{s}}{m_t^2} \right)\right]\right\},
\end{align}
where $N_c$(=3) is the color factor. The full expression of the vertex function
$\Gamma_{hhh}^\text{SM}(p_1^2,p_2^2,p_3^2)$ is also given in Appendix~A for completeness.
In numerical analysis, we include the SM one-loop correction to the $hhh$ coupling constant.
}
with $v$ ($\simeq 246$ GeV) being the VEV and $m_h$ being the mass of the Higgs boson $h$.
This constant shift can be realized when there is the dimension six operator in the Higgs potential\cite{Ref:hhhPLC,dim6}. Quantum corrections to the $hhh$ coupling constant due to
the bosonic loop can also provide the constant shift of the $hhh$ coupling constant
approximately~\cite{Ref:KOSY}.

\begin{figure}[tb]
\includegraphics[height=2cm]{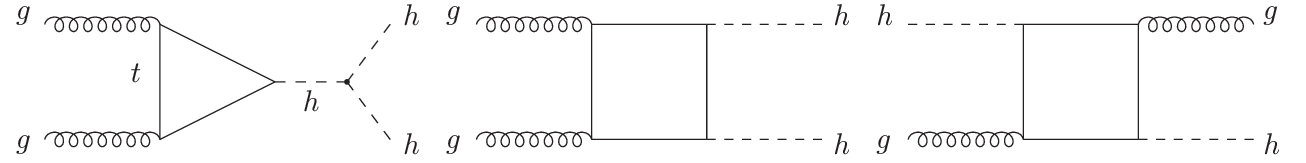}
\caption{The double Higgs boson production process $gg\to hh$ via gluon fusion at the hadron
collider.}
\label{FIG:diag-gghh}
\end{figure}

At the LHC, the largest cross section of the Higgs boson pair production comes from
the gluon fusion mechanism~\cite{Ref:gghh-qqhh,Ref:gghh,Ref:gghh-Vhh-gghh}.
Feynman diagrams for $gg\to hh$ are depicted in FIG.~\ref{FIG:diag-gghh}.
The triangular loop diagrams contain information of the $hhh$ coupling constant.
The parton level cross sections are calculated at the leading order as~\cite{Ref:gghh}
\begin{align}
{\widehat \sigma}(gg\to hh)
&=\int_{{\hat t}_-}^{{\hat t}_+}d{\hat t}\,
\frac1{2^2}\frac1{8^2}\frac1{2!}\frac1{16\pi{\hat s}^2}\frac{2\alpha_S^2}{(4\pi)^2}
\left\{\left|\frac{\lambda_{hhh}\,v}{{\hat s}-m_h^2}F_\triangle+F_\Box\right|^2
+\left|G_\Box\right|^2\right\},
\end{align}
where $F_\triangle$ is the loop function for the triangular diagram, while $F_\Box$
and $G_\Box$ are those for box diagrams which, respectively, correspond to the invariant
amplitudes for same and opposite polarizations of incoming
gluons~\cite{Ref:hhh-sensitivity}: see Appendix~B.
The invariant mass distribution can be obtained by multiplying the gluon-gluon
luminosity function as
\begin{align}
\frac{d\sigma(gg\to hh)}{dM_{hh}}
&=\frac{2M_{hh}}{s}{\widehat \sigma}(gg\to hh)
\frac{dL_{gg}}{d\tau},
\end{align}
where $M_{hh}=\sqrt{\hat s}$, $\tau={\hat s}/s$, and
\begin{align}
\frac{dL_{gg}}{d\tau}
&=\int_\tau^1 \frac{dx}{x}f_g(x,\mu_F^{}=M_{hh})f_g(\tau/x,\mu_F^{}=M_{hh}),
\end{align}
where $f_g(x,\mu_F^{})$ is the parton distribution function of gluons. In our numerical
calculation, the CTEQ6L parton distribution function is used~\cite{Ref:cteq6}. The loop
integrals are evaluated by a package; LoopTools~\cite{Ref:LoopTools}.

It is well known that this process receives large QCD corrections
\footnote{
The next-to-leading order (NLO) QCD corrections to this process have been computed
in the heavy top-quark mass limit in Ref.~\cite{Ref:NLOGGhh}, which give an over
all factor $K\simeq 1.9$ ($K$-factor) for $\mu_F^{}=M_{hh}$. The smaller value of
$K\simeq 1.65$ for $\mu_F=m_h$ was suggested by Ref.~\cite{Ref:hhh-sensitivity}.
The correction mainly comes from the initial state radiation of gluons. It is known
that this kind of approximation works well in the single Higgs production via the
gluon fusion mechanism, where the NLO cross section is evaluated by the leading order
$gg\to h$ cross section for a finite top-quark mass with the $K$-factor in the large
$m_t$ limit. The running of strong coupling constant can also change the cross
section by $25$--$50$\%~\cite{Ref:hhh-sensitivity}.
}.
Although the NLO calculation is very important in evaluating this process, throughout
this paper we totally neglect NLO QCD corrections in our calculations of the cross
section in various new physics models. The QCD corrections in each new physics model
are currently unknown so that the computation of these corrections is beyond the scope
of this paper.

\begin{figure}[tb]
\includegraphics[height=5.5cm]{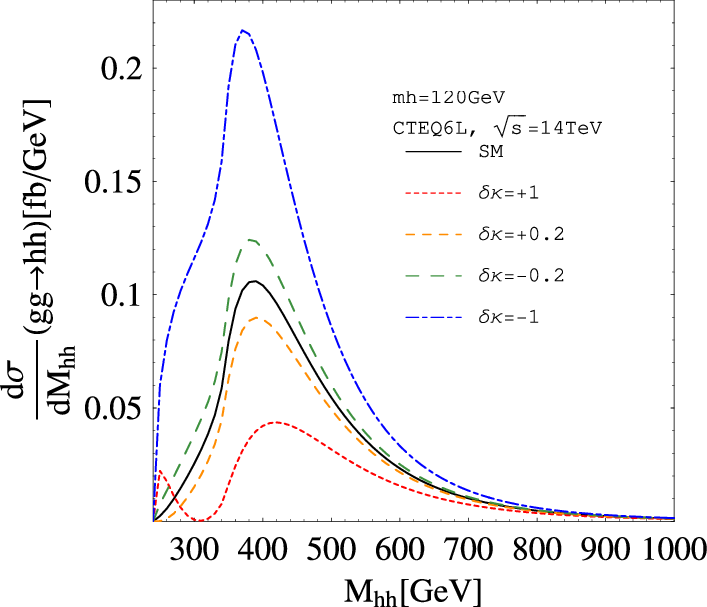}
\includegraphics[height=5.5cm]{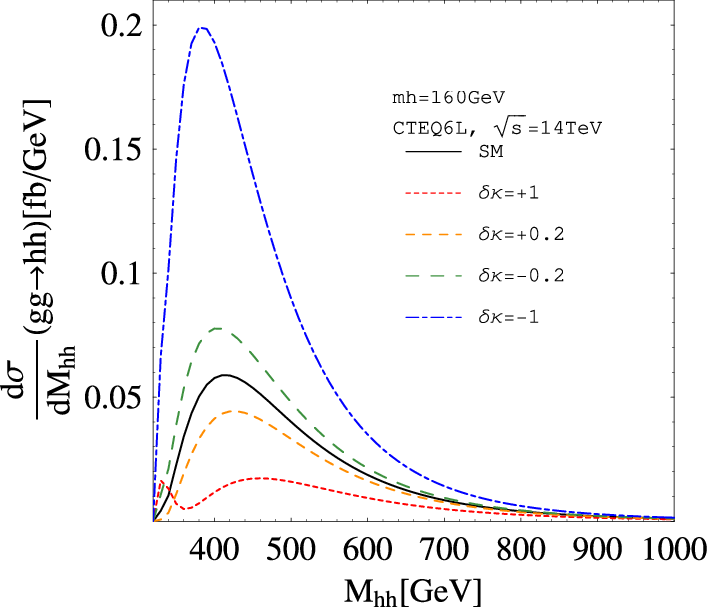}
\caption{The invariant mass distribution of the cross section of $gg\to hh$ process
at the LHC with $\sqrt{s}=14$ TeV for $m_h=120$ GeV (left) and $m_h=160$ GeV (right).
The solid, dotted, dashed, long-dashed and dot-dashed curved lines denote the SM prediction,
the SM with the positive $100\%$ correction to the $hhh$ coupling constant, that with
the $+20\%$ correction, that with the $-20\%$ correction, and that with the $-100\%$
correction, respectively.}
\label{FIG:GGhhinvSM}
\end{figure}
In FIG.~\ref{FIG:GGhhinvSM}, we show the invariant mass distributions of the cross
section of $gg\to hh$ process with the deviation of the $hhh$ coupling constant for
$m_h=120$ GeV (left) and for $m_h=160$ GeV (right). Throughout this paper we take
the top-quark mass to be $171.2$ GeV. These solid, dotted/dashed and long-dashed/dot-dashed
curves represent the SM prediction including the SM one-loop effect on the $hhh$ coupling
constant, that with constructive deviations $\Delta\kappa=+1.0$ and $+0.2$ and that with
destructive deviations $\Delta\kappa=-1.0$ and $-0.2$, respectively. The total cross section
is about $20$ $(10)$ fb for $m_h=120$ $(160)$ GeV in the SM. Only for $\Delta\kappa=+1.0$,
a small peak comes from the large $hhh$ coupling constant through the triangular diagram
in the near threshold region. The peaks can be found around $M_{hh}\sim 400$ GeV, which are
caused by the interference effect of the triangular and the box diagrams. Since these two
contributions are destructive to each other, the positive (negative) variation of the $hhh$
coupling constant makes the cross sections small (large) in this process. This means that
in the $gg\to hh$ process the sensitivity is getting better for the negative contribution
to the $hhh$ coupling constant and vice versa. If we have additional colored particles
in the new physics model, this situation could be changed.

\begin{figure}[tb]
\includegraphics[height=2.5cm]{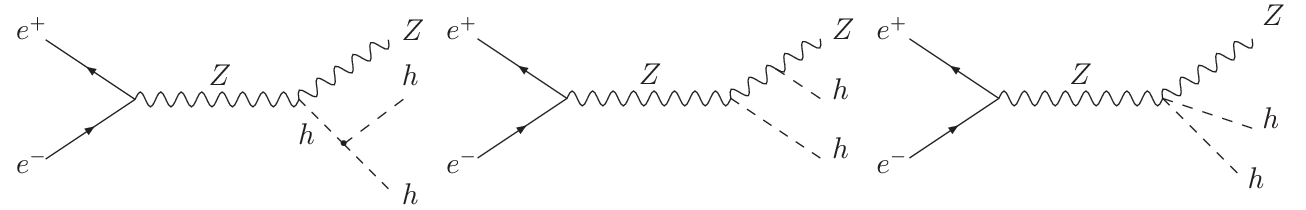}
\includegraphics[height=4cm]{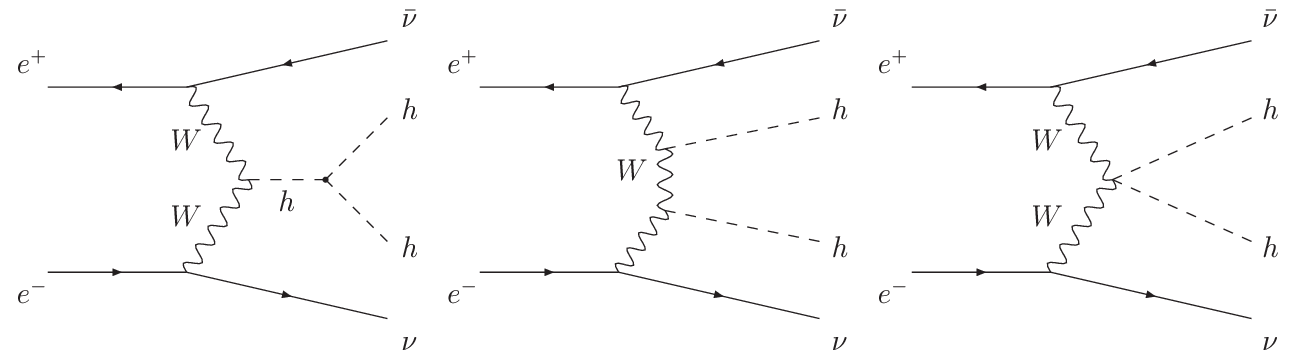}
\caption{The double Higgs boson production at the $e^+e^-$ collider.
The double-Higgs-strahlung process $e^+e^-\to hhZ$ and the vector boson fusion process
$e^+e^- \to hh\nu_e \bar\nu_e$.}
\label{FIG:diag-eehhz-wfusion}
\end{figure}

At an electron-positron linear collider, the $hhh$ coupling constant will be measured by
the double-Higgs-strahlung~\cite{Ref:hhZ} and the Higgs boson pair production via the
$W$ boson fusion mechanism~\cite{Ref:hhnunu}. Feynman diagrams for these processes are shown
in FIG.~\ref{FIG:diag-eehhz-wfusion}. The $e^+e^-\to hhZ$ process may be a promising channel
at the ILC to measure the $hhh$ coupling constant for light Higgs bosons because of the simple
kinematical structure. Since relatively larger collision energy is required for three body
final states of $hhZ$, the $s$-channel nature of the process may decrease the cross section.
On the other hand, if we have large enough energy, one can control the collision energy to
obtain the maximal production rate.
\begin{figure}[tb]
\includegraphics[height=5.5cm]{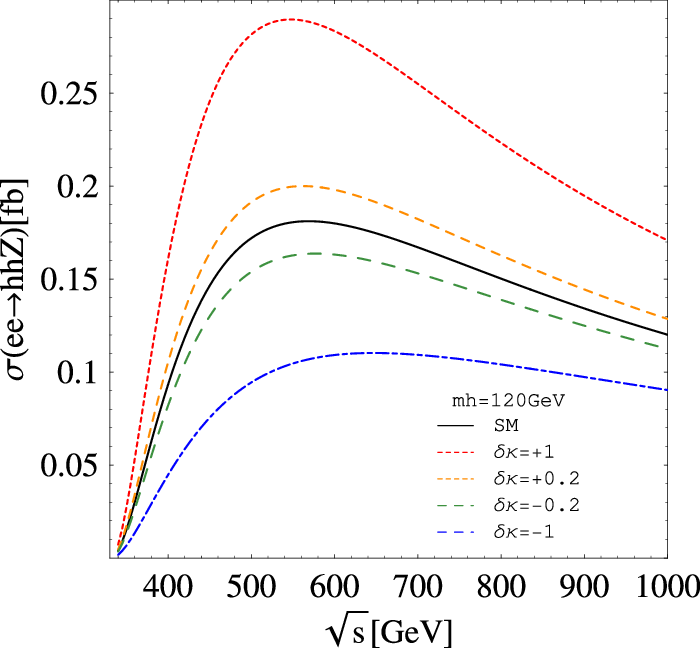}
\includegraphics[height=5.5cm]{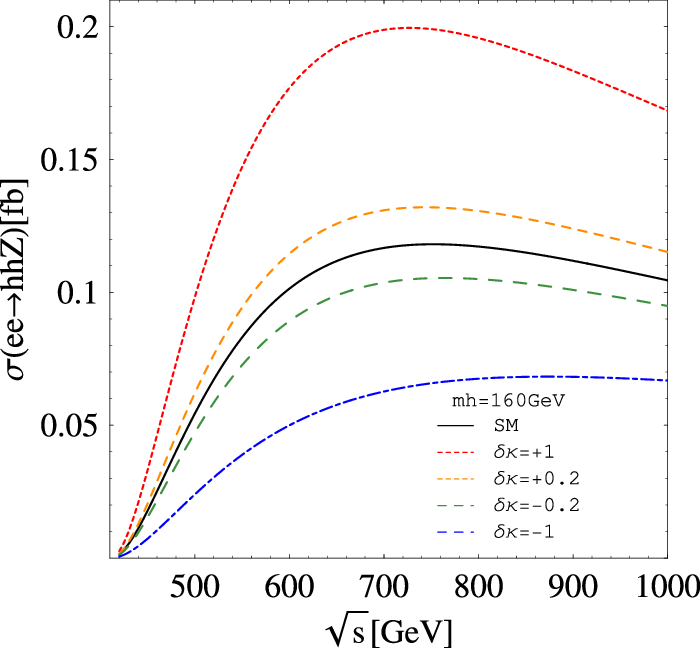}
\caption{The cross sections of $e^+e^-\to hhZ$ process at the ILC as a function of collision
energy $\sqrt{s}$ for $m_h=120$ GeV (left) and $m_h=160$ GeV (right).
}
\label{FIG:eehhZSM}
\end{figure}
In FIG.~\ref{FIG:eehhZSM}, the cross sections of the double-Higgs-strahlung are evaluated as
a function of $e^+e^-$ center of mass energy $\sqrt{s}$. The left (right) panel shows the case
with the Higgs boson mass to be $m_h=120 (160)$ GeV. The curves are presented in the same manner
as in FIG.~\ref{FIG:GGhhinvSM}. Under the variation of the $hhh$ coupling constant, the cross
section of the double-Higgs-strahlung has the opposite correlation to that of $gg\to hh$.
Therefore, the positive contributions to the $hhh$ coupling constant has an advantage to obtain
better sensitivities.

\begin{figure}[tb]
\includegraphics[height=8cm]{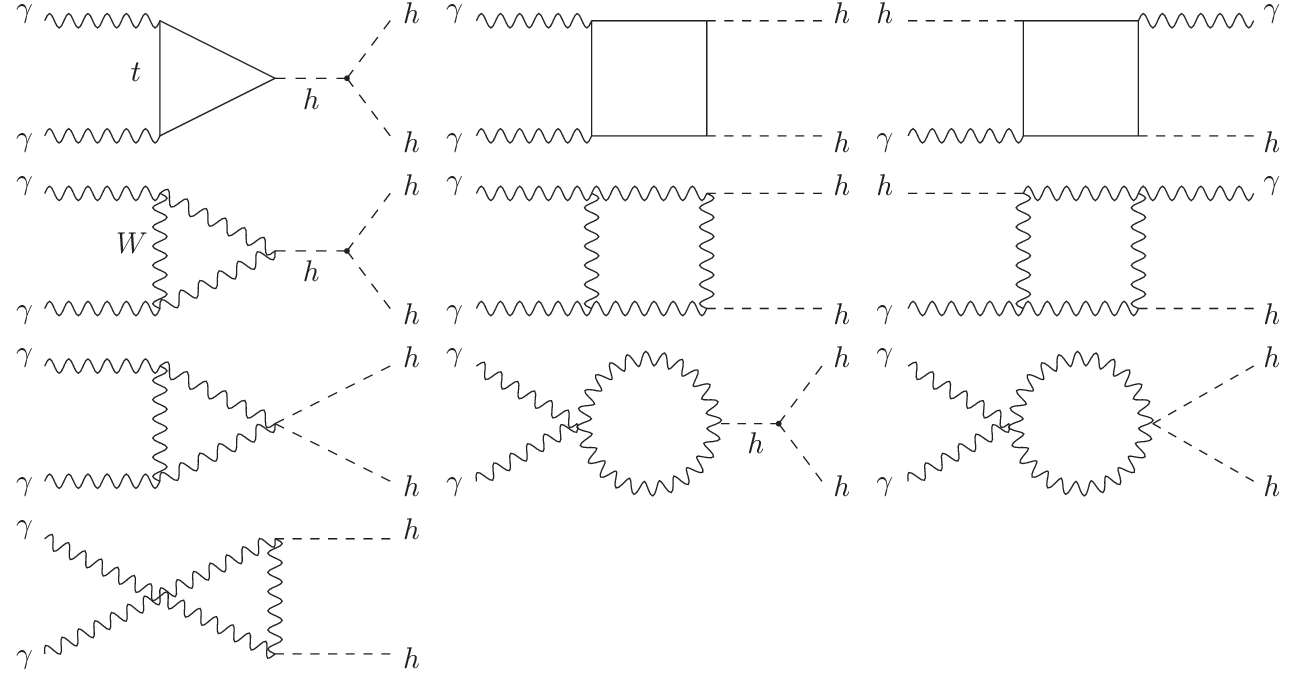}
\caption{The double Higgs boson production process $\gamma\gamma\to hh$ at the photon collider.}
\label{FIG:diag-gamgamhh}
\end{figure}

At a high energy lepton collider, the hard photons can be obtained from the Compton back
scattering method~\cite{Ref:PLC}. By using hard photons, Higgs boson pairs can be produced
in $\gamma\gamma\to hh$ process. Feynman diagrams for this process are shown in
FIG.~\ref{FIG:diag-gamgamhh}, and the helicity specified cross sections are given by
\begin{align}
{\hat \sigma}^{\lambda_1\lambda_2}\equiv
{\widehat \sigma}(\gamma_{\lambda_1}\gamma_{\lambda_2}\to hh)
&=\int_{{\hat t}_-}^{{\hat t}_+}d{\hat t}\,
\frac1{2!}\frac1{16\pi{\hat s}^2}\frac{\alpha_\text{EM}^2}{(4\pi)^2}
\left|\frac{\lambda_{hhh}\,v}{{\hat s}-m_h^2}H_\triangle^{\lambda_1\lambda_2}
+H_\Box^{\lambda_1\lambda_2}\right|^2,
\end{align}
where $H_\triangle^{\lambda_1\lambda_2}$ and $H_\Box^{\lambda_1\lambda_2}$ are the loop
functions~\cite{Ref:hhhPLC}(see Appendix B).
The total cross section is calculated by convoluting with the photon luminosity function
$f_\gamma(y, x)$, where $x=4E_e\omega_0/m_e^2$ can be controlled by the frequency $\omega_0$
of the laser photon, as
\begin{align}
\sigma_{\gamma\gamma\to hh}^{ee}
&=\int_{\tau_{hh}^{}}^{y_m^2}d\tau\int_\tau^{y_m}\frac{dy}{y}
\left[\frac{1+\xi^\gamma_1\xi^\gamma_2}2{\hat \sigma}^{++}
+\frac{1-\xi^\gamma_1\xi^\gamma_2}2{\hat \sigma}^{+-}\right]
f_\gamma(y, x)f_\gamma(\tau/y, x),
\end{align}
where $\xi^\gamma$ is the mean helicity of the photon. The photon luminosity spectrum
is given by
\begin{align}
f_\gamma(y, x)
&= \frac1{D(x)}\left[\frac1{1-y}+1-y-4r(1-r)-2\lambda^e \lambda^\gamma r x (2r-1)(2-y)\right],\\
D(x)&=\left(1-\frac4x-\frac8{x^2}\right)\ln(1+x)+\frac12+\frac8x-\frac1{2(1+x)^2}\nonumber\\
&\quad+2\lambda^e\lambda^\gamma\left[(1+\frac2x)\ln(1+x)-\frac52+\frac1{1+x}-\frac1{2(1+x)^2}\right],
\end{align}
where $r=\frac{y}{x(1-y)}$ and $\lambda^e (\lambda^\gamma)$ is the helicity of the incident
electron (photon)~\cite{Ref:PLC}. The maximal energy fraction of photon $y_m=\frac{x}{1+x}$
is fixed by the kinematics of the Compton scattering at the photon collider.
\begin{figure}[tb]
\includegraphics[height=5.5cm]{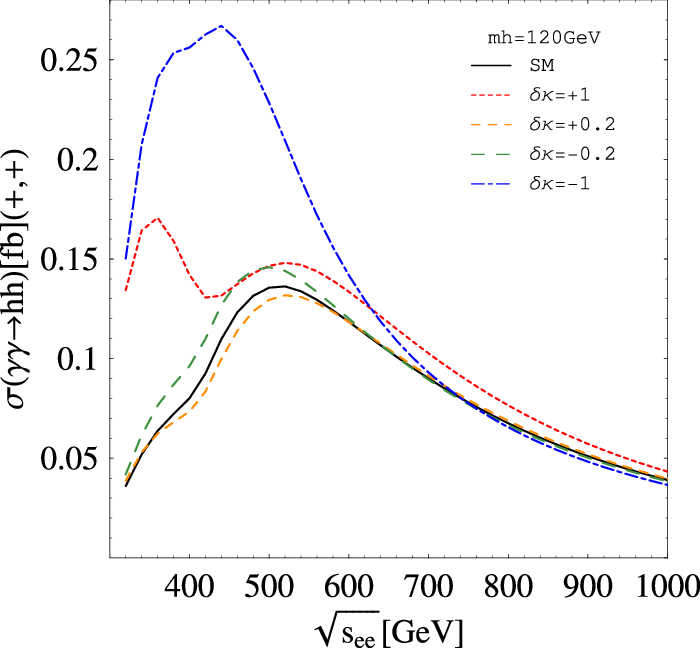}
\includegraphics[height=5.5cm]{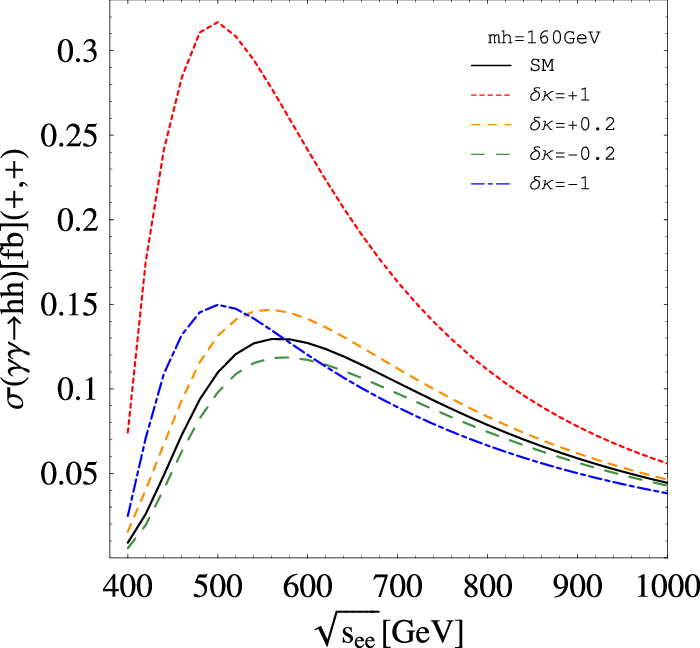}
\caption{The full cross section of $e^-e^-\left(\gamma(+)\gamma(+)\right)\to hh$ process
as a function of $\sqrt{s}_{ee}$ for $m_h=120$ GeV (left) and $m_h=160$ GeV (right).
}
\label{FIG:AAhhSM}
\end{figure}
In FIG.~\ref{FIG:AAhhSM}, the full cross sections of
$e^-e^-\left(\gamma(+)\gamma(+)\right)\to hh$ are shown as a function of the energy of
the $e^-e^-$ system. We here choose the same sign polarizations for initial photons in order
to efficiently extract information of the $hhh$ coupling constant. The parameter $x$ is taken
to be $4.8$, which can be tuned by the frequency of the laser photon. The curves are given
in the same manner as in FIG.~\ref{FIG:GGhhinvSM}.
The situation is very different from $gg\to hh$ at the LHC. Energies of initial gluons are
widely varied at a hadron collider, while back-scattered photons at the PLC have narrow band
spectra. Therefore, we can tune the effective energy of photons at the PLC to some extent.
The relative strength of the $W$ boson and the top-quark loop diagrams strongly depends on
the collision energy and the Higgs boson mass. Only for $m_h=120$ GeV, the large $hhh$ coupling
constant case ($\Delta\kappa=+1.0$) shows a peak at the near threshold regime. It is found that
the negative deviation of the $hhh$ coupling constant makes cross section large for $m_h=120$ GeV
(left), while it has an opposite effect on the cross section for $m_h=160$ GeV (right).

\begin{figure}[tb]
\includegraphics[height=5.5cm]{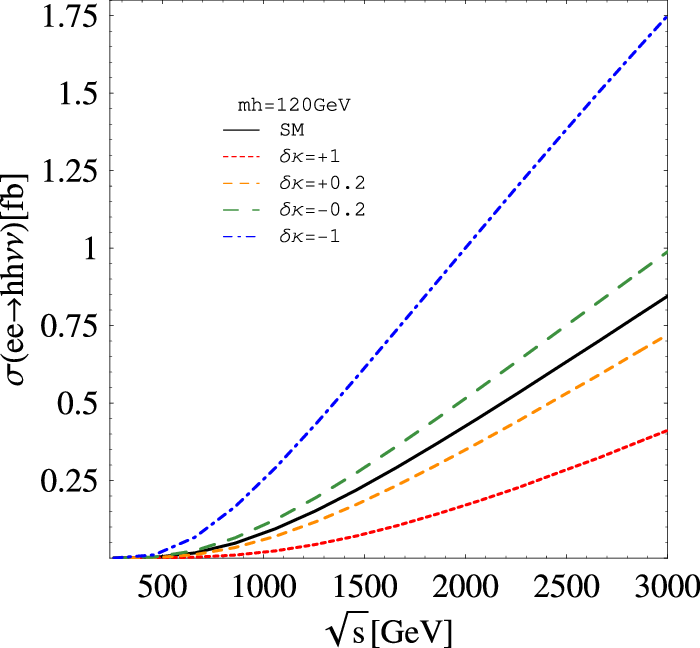}
\includegraphics[height=5.5cm]{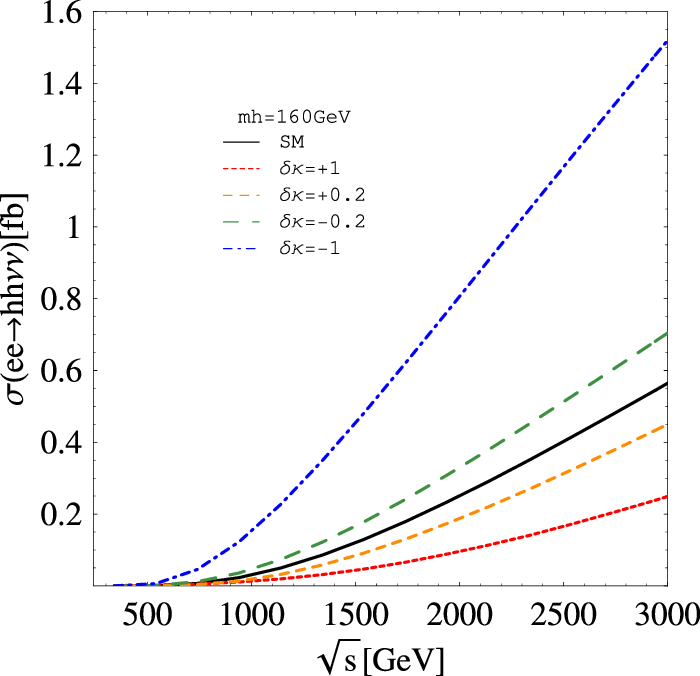}
\caption{The cross sections of $e^+e^-\to hh\nu\bar{\nu}$ process at the ILC as a function of
collision energy $\sqrt{s}$ for $m_h=120$ GeV (left) and $m_h=160$ GeV (right).}
\label{FIG:eehhNNSM}
\end{figure}
If we go to further high energy $e^+e^-$ colliders, the second stage of the ILC or the CLIC,
the Higgs boson pair production via the $W$ boson fusion mechanism becomes important~\cite{Ref:hhnunu}.
The cross section increases for higher energy because of the $t$-channel enhancement of $W^+W^-\to hh$
subprocess. In FIG.~\ref{FIG:eehhNNSM}, we evaluate the production rate for $e^+e^-\to hh\nu\bar\nu$
by CalcHEP~\cite{Ref:CalcHEP}. For both $m_h=120$ GeV (left) and $m_h=160$ GeV (right) cases,
the cross section simply scales as a function of energy and can be much larger than those of
$e^+e^-\to hhZ$ and $\gamma\gamma\to hh$. The $\Delta \kappa$ dependence in the cross section of
$e^+e^-\to hh\nu\bar\nu$ is opposite to that in $e^+e^-\to hhZ$; i.e., a larger cross section for
$e^+e^-\to hh\nu\bar\nu$ is obtained for a larger $|\Delta \kappa|$ value with a negative sign.

\section{Comparison of the Higgs boson pair creation processes in different models}
In this section, we study cross sections for the double Higgs boson production processes
$gg\to hh$~\cite{Ref:gghh-qqhh,Ref:gghh,Ref:gghh-Vhh-gghh}, $e^+e^-\to hhZ$~\cite{Ref:hhZ},
$e^+e^- \to hh\nu \bar \nu$~\cite{Ref:hhnunu} and $\gamma\gamma\to hh$~\cite{Ref:hhhPLC}
in four different models; i.e., the THDM, the model with scalar leptoquarks, that with
chiral fourth generation quarks and leptons, and that with vectorlike quarks.
In these models, one-loop contributions to the $hhh$ coupling constant can be
nondecoupling even when new particles are heavier than the electroweak scale, so that
large deviations in the $hhh$ coupling constant can affect the cross sections of these
double Higgs boson production processes. Unlike the analysis with the constant shift with
$\Delta \kappa$ in the previous section, the energy dependencies in the $hhh$ vertex function
are also included in our evaluation here. Furthermore, for the one-loop induced processes
such as $gg\to hh$ and $\gamma\gamma\to hh$, the contribution of additional one-loop diagrams
where the new particles are running in the loop can be significant.

\subsection{Two-Higgs-doublet model}
The THDM~\cite{Ref:HHG} is the simplest extension of the Higgs sector in the SM, which can appear
in various new physics scenarios such as the minimal supersymmetric SM~\cite{Ref:SUSY},
the top color model~\cite{Ref:topcolor}, radiative seesaw models for neutrinos~\cite{1OPT2,Ref:Zee}
and models of electroweak baryogenesis~\cite{Ref:thdmEWBG}. The Higgs scalar doublets interact
with other fields purely by the electroweak force. Therefore, in addition to the change in
the cross section due to the quantum correction to the $hhh$ coupling constant, the contribution
of the charged Higgs boson loop can affect the cross section of $\gamma\gamma\to hh$, while
the loop effect of extra scalar bosons appear only through the correction to the $hhh$ coupling
constant for $gg\to hh$, $e^+e^-\to hhZ$ and $e^+e^-\to hh\nu\bar \nu$.

The potential of the THDM with a softly-broken discrete $Z_2$ symmetry is given by
\begin{align}
V_\text{THDM}
&= m_1^2\Phi_1^\dag\Phi_1+m_2^2\Phi_2^\dag\Phi_2
-\left(m_3^2\Phi_1^\dag\Phi_2+\text{H.c.}\right)
+\frac{\lambda_1}2(\Phi_1^\dag\Phi_1)^2
+\frac{\lambda_2}2(\Phi_2^\dag\Phi_2)^2\nonumber \\
&\qquad+\lambda_3(\Phi_1^\dag\Phi_1)(\Phi_2^\dag\Phi_2)
+\lambda_4(\Phi_1^\dag\Phi_2)(\Phi_2^\dag\Phi_1)
+\left[\frac{\lambda_5}2(\Phi_1^\dag\Phi_2)^2+\text{H.c.}\right],\label{Eq:THDMPot}
\end{align}
where $\Phi_i$ $(i=1,2)$ are scalar isospin doublet fields with the hypercharge of $+1/2$, which
transform as $\Phi_1 \to \Phi_1$ and $\Phi_2 \to -\Phi_2$ under the $Z_2$. Although $m_3^2$ and
$\lambda_5$ are complex in general, we here take them to be real assuming the CP invariance.
The two Higgs doublet fields can then be parameterized as
\begin{align}
\Phi_i=\begin{pmatrix}
\omega^+_i\\\frac1{\sqrt2}(v_i+h_i+i\,z_i)
\end{pmatrix}.
\end{align}
There are $8$ degrees of freedom in the two complex scalar doublet fields. Three of them are
absorbed as the longitudinal components of the weak gauge bosons. The remaining five convert into
the mass eigenstates, two CP even Higgs bosons $(h,H)$, a CP odd Higgs boson $(A)$, and a pair of
charged Higgs bosons $(H^\pm)$. The eight parameters $m_1^2$--$m_3^2$ and $\lambda_1$--$\lambda_5$
are replaced by the VEV $v$, the mixing angle of CP even Higgs bosons $\alpha$, the ratio of
VEVs $\tan\beta=v_2/v_1$, the Higgs boson masses $m_h^{},m_H^{},m_A^{},m_{H^\pm}^{}$ and the soft
breaking parameter $M^2=m_3^2/(\sin\beta\cos\beta)$.
The parameters in the Higgs potential can be constrained by imposing theoretical assumptions
such as perturbative unitarity~\cite{Ref:PUsm,Ref:PUthdm} and vacuum stability~\cite{Ref:VSthdm}.
If we require stability of the theory below a given cutoff scale $\Lambda$ imposing the conditions
of vacuum stability and triviality, the Higgs boson parameters are constrained as a function of
$\Lambda$ by the renormalization group equation analysis\cite{Ref:TBthdm}.

Here, we consider the ``SM-like'' case with $\sin(\beta-\alpha)=1$ where only the lighter CP-even
Higgs boson $h$ couples to the weak gauge boson as $VVh$~\cite{Ref:GunionHaber}. In this case,
all the coupling constants of $h$ to the SM particles take the same form as those in the SM at
the tree level. The $hhh$ coupling constant is also described by the same tree-level formula as
in the SM. The difference appears at the loop level due to the one-loop contribution of the extra
scalars.

Under the imposed softly-broken discrete symmetry, there can be four types of Yukawa
interactions~\cite{Ref:Yukawathdm1,Ref:Yukawathdm2}. Although in general there can be large
phenomenological differences among the different types of Yukawa interaction, especially in flavor
physics, we do not specify the type of Yukawa interaction in this paper, because there is no proper
difference in the discussion here in the SM-like limit where $h$ behaves as if it were the SM Higgs
boson at the tree level
\footnote{
The allowed regions of $m_{H^\pm}$ and $\tan\beta$ can receive constraint from the flavor physics
data such as $b\to s \gamma$ depending on the type of Yukawa interaction~\cite{Ref:bsTHDM,Ref:Yukawathdm1}. }.
%

The extra Higgs bosons have been searched at the LEP experiment. The lower mass bound for the 
CP-even Higgs boson is $m_H^{}>92.8$ GeV, and that for the CP-odd Higgs boson is $m_A^{}>93.4$ GeV 
in the minimal supersymmetric SM whose Higgs sector is the THDM~\cite{Ref:LEP}. The bound for 
charged Higgs boson mass has also been set as $m_{H^{\pm}}^{}>79.3$ GeV~\cite{Ref:LEP}.
%
The electroweak precision data from the LEP experiment may indicate that the Higgs sector
approximately respects the custodial $SU(2)$ symmetry~\cite{Ref:Custodial}. This symmetry becomes
exact in the Higgs potential in the limit of $m_A^{}=m_{H^\pm}^{}$ with arbitrary $\sin(\beta-\alpha)$
or in the limit of $m_H^{}=m_{H^\pm}^{}$ with $\sin(\beta-\alpha)=1$ (or $m_h^{}=m_{H^\pm}^{}$ with
$\cos(\beta-\alpha)=1$). The breaking of the custodial symmetry in the two-Higgs-doublet potential
gives large contribution to the $\widehat{T}$ parameter which is proportional to the mass differences
of extra Higgs bosons. In order to suppress these contributions, we take their masses to be degenerate
in the following discussion; i.e., $m_H=m_A=m_{H^\pm}$.

The one-loop correction to the $hhh$ coupling constant in the THDM is evaluated as~\cite{Ref:KOSY}
\begin{figure}
\includegraphics[height=5.5cm]{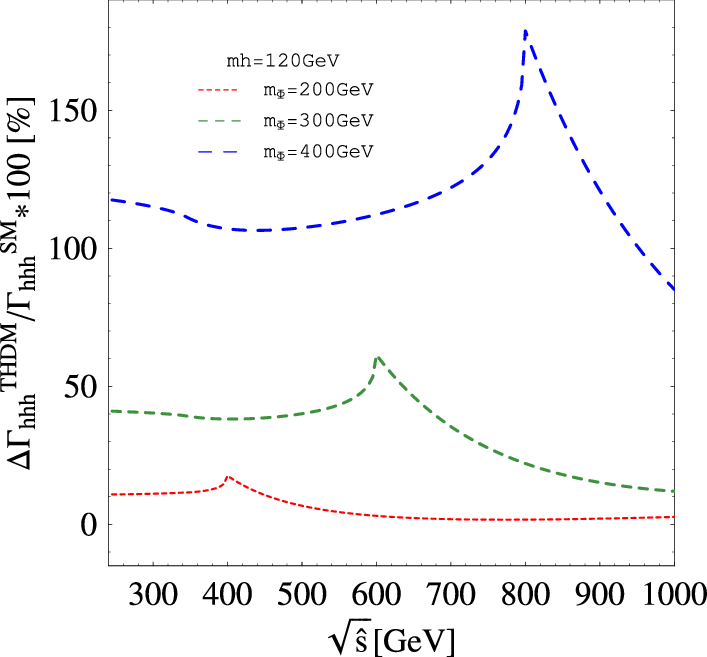}
\includegraphics[height=5.5cm]{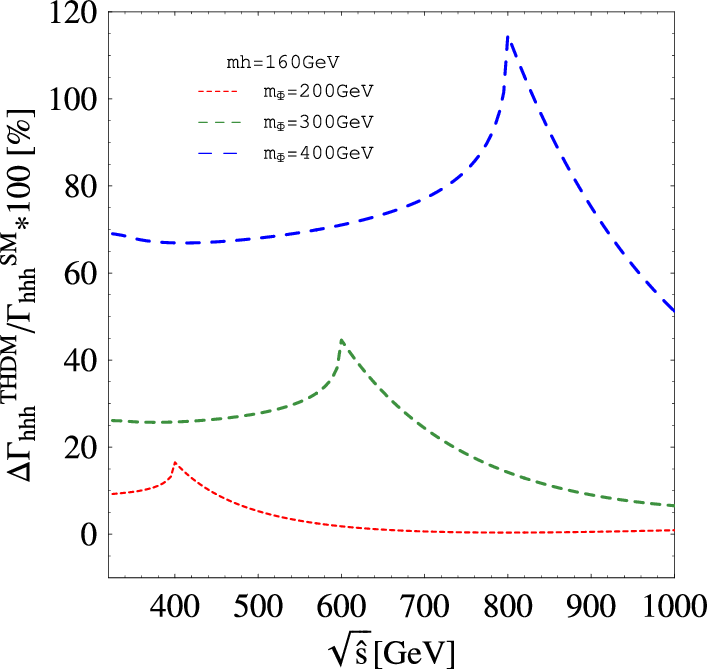}
\caption{The rates for one-loop contributions from $H, A, H^\pm$ in the THDM to the $hhh$ coupling
constant for $m_h=120$~GeV (left) and for $m_h=160$~GeV (right).}
\label{FIG:hhh-thdm}
\end{figure}
\begin{align}
\frac{\Gamma_{hhh}^\text{THDM}}{\Gamma_{hhh}^\text{SM}}
&\simeq 1+\frac{m_{H^\pm}^4}{6\pi^2v^2m_h^2}
\left(1-\frac{M^2}{m_{H^\pm}^2}\right)^3
+\frac{m_{H}^4}{12\pi^2v^2m_h^2}\left(1-\frac{M^2}{m_{H}^2}\right)^3
+\frac{m_{A}^4}{12\pi^2v^2m_h^2}\left(1-\frac{M^2}{m_{A}^2}\right)^3,
\end{align}
where $\sin(\beta-\alpha)=1$ is taken.
The deviation from the SM results can be very large when $M^2\simeq 0$. The full calculation of
the vertex function is shown in Ref.~\cite{Ref:KOSY}
\footnote{
In the case other than $\sin(\beta-\alpha)=1$, the $hhh$ coupling constant can deviate from
the SM value at the tree level because of the mixing between $h$ and $H$. The discussion at
the one-loop level with including such a mixing effect is also given in Ref.~\cite{Ref:KOSY}}.
In FIG.~\ref{FIG:hhh-thdm}, the deviation in the effective $hhh$ coupling constant from the SM
value is shown as a function of $\sqrt{\hat{s}}$, the energy of $h^\ast \to hh$ 
in the THDM~\cite{Ref:KOSY}. The mass of the SM-like Higgs boson is taken to be $m_h=120$ GeV
(left) and $m_h=160$ GeV (right). The masses of extra Higgs bosons are taken to be
$m_\Phi^{}=200$ GeV (dotted line), $m_\Phi^{}=300$ GeV (dashed line), and $m_\Phi^{}=400$ GeV 
(long-dashed line), where $m_\Phi^{} \equiv m_H^{}=m_A^{}=m_{H^\pm}^{}$. These effects can be 
about $120$--$70$\% for $m_h=120$--$160$ GeV with $m_\Phi^{}\sim 400$ GeV.
FIG.~\ref{FIG:hhh-thdm} shows that the deviation in the $hhh$ coupling constant can be
approximately described by the analysis with a constant shift by the factor of $(1+\Delta\kappa)$.

\begin{figure}
\includegraphics[height=5.5cm]{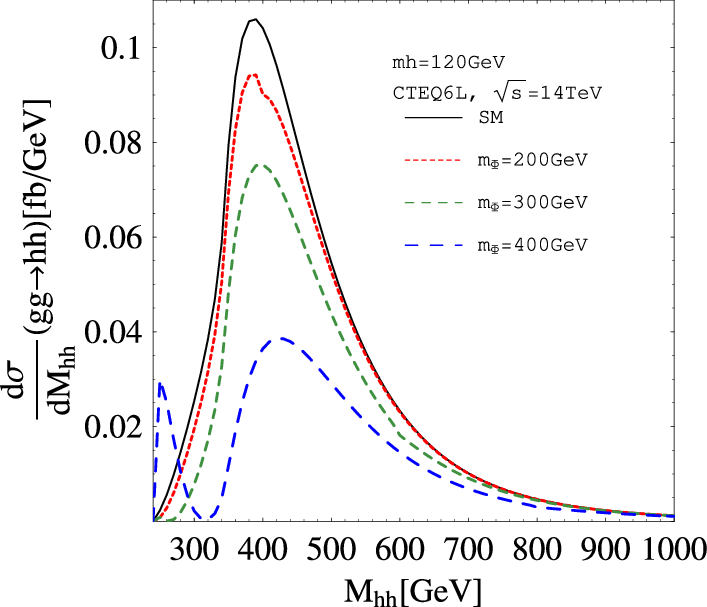}
\includegraphics[height=5.5cm]{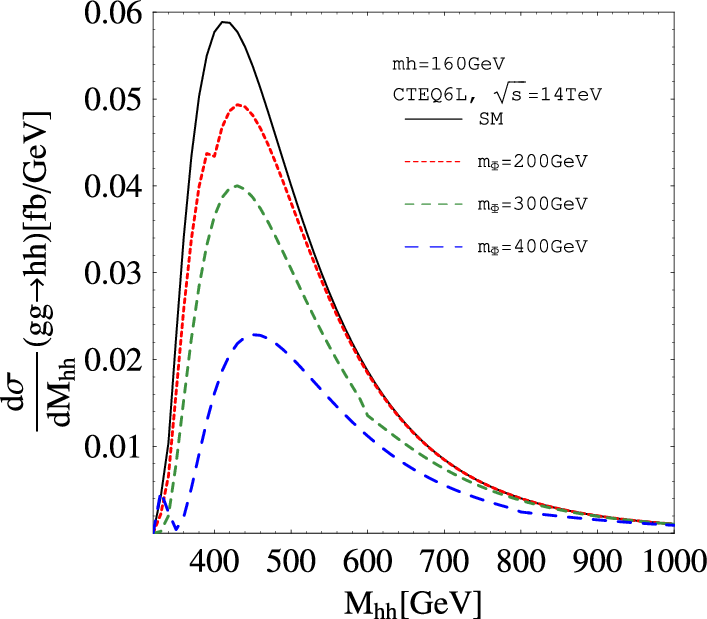}
\caption{The invariant mass distribution of $gg\to hh$ process at the LHC with $\sqrt{s}=14$ TeV
for $m_h=120$ GeV (left) and $m_h=160$ GeV (right) in the THDM.}
\label{FIG:GGhh-inv-thdm}
\end{figure}
In FIG.~\ref{FIG:GGhh-inv-thdm}, the invariant mass distribution of the differential cross section
for $gg\to hh$ at the LHC is shown in the THDM
~\footnote{
Cases without $\sin(\beta-\alpha)=1$ were considered in Ref.~\cite{Ref:ggArhrib}.
The $s$-channel resonance effect of new physics particles were also discussed in Ref.~\cite{Ref:f6}.}.
The curves are given in the same manner as in FIG.~\ref{FIG:hhh-thdm}, and the SM predictions are
also denoted by solid curves for comparison. Higgs bosons do not couple to gluons at the tree level,
so that the one-loop effect of the extra Higgs bosons only appear in the correction to the $hhh$
coupling constant. For larger extra scalar masses $m_\Phi= 400$ GeV, peaks can be found in the near
threshold region of a Higgs pair, which come from the enhancement of the $hhh$ coupling constant.
There are also peaks around $M_{hh}\sim 400$ GeV, which are interference effects between
the triangular and the box diagrams. Those contributions weaken each other, and hence the enhancement
of the $hhh$ coupling constant decreases the cross section as in the SM with constant deviation.

\begin{figure}
\includegraphics[height=5.5cm]{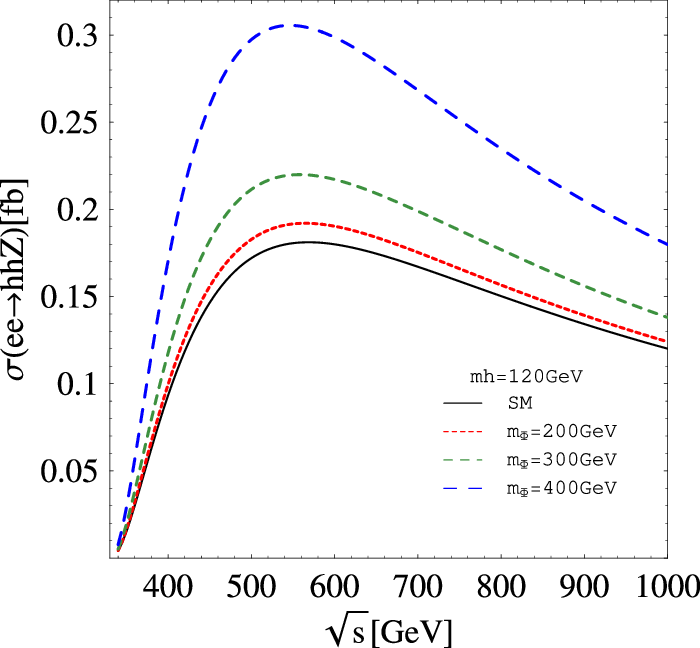}
\includegraphics[height=5.5cm]{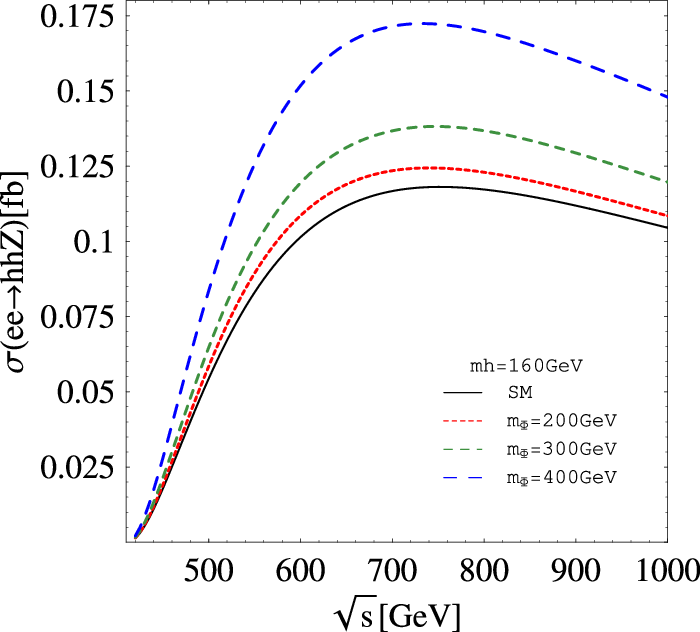}
\caption{The cross section of $e^+e^-\to hhZ$ process as a function of $\sqrt{s}$ for $m_h=120$ GeV
(left) and $m_h=160$ GeV (right) in the THDM.}
\label{FIG:eehhZ-thdm}
\end{figure}
In FIG.~\ref{FIG:eehhZ-thdm}, we show the cross section of the process $e^+e^-\to hhZ$ as a function
of the collision energy $\sqrt{s}$ in the THDM
~\footnote{
More general types of double Higgs-boson production processes at $e^+e^- $ colliders were studied 
in Ref.~\cite{Ref:f7}.}. 
The curves are presented in the same manner as in FIG.~\ref{FIG:GGhh-inv-thdm}. Relatively large nondecoupling effect of the extra scalar bosons can
appear in the radiative correction to the $hhh$ coupling constant.

\begin{figure}
\includegraphics[height=5.5cm]{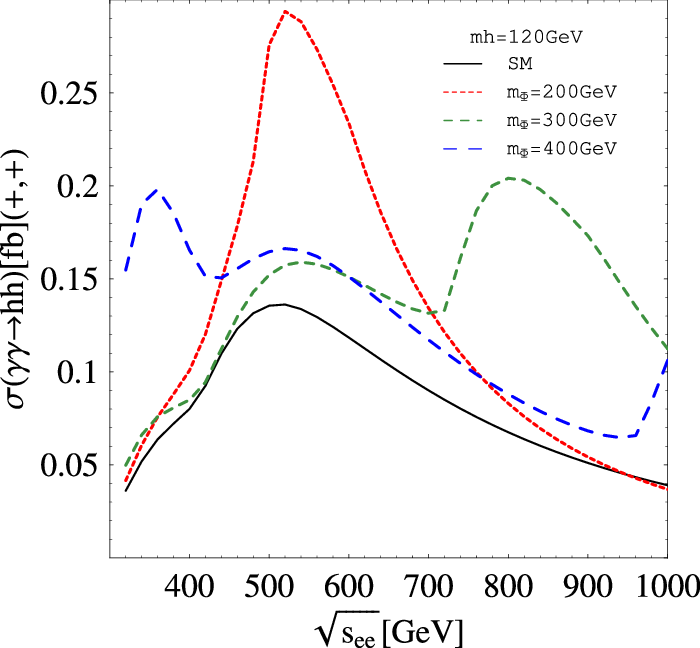}
\includegraphics[height=5.5cm]{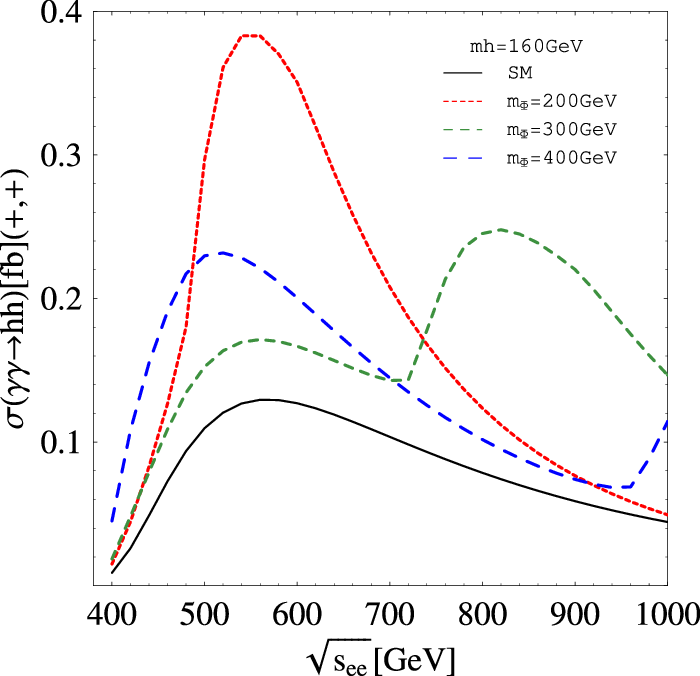}
\caption{The cross section of $\gamma(+)\gamma(+)\to hh$ process at the photon collider option at
the ILC as a function of the $e^-e^-$ collision energy for $m_h=120$ GeV (left) and $m_h=160$ GeV
(right) in the THDM.}
\label{FIG:GamGamhh-thdm}
\end{figure}
In FIG.~\ref{FIG:GamGamhh-thdm}, the cross sections of the Higgs pair production at the PLC are
given for the THDM~\cite{Ref:Hollik,Ref:AHKOT,Ref:Arhrib}. The extra Higgs boson can contribute to
the corrections of the $hhh$ coupling constant as well as $\gamma\gamma\to hh$ process.
The $hhh$ coupling constant can be probed by choosing the collision energy to be near threshold
region for relatively heavy extra Higgs bosons $m_\Phi^{}\gtrsim 400$ GeV. There are threshold
enhancement from the box diagrams after $\sqrt{s_{ee}}\sim 2m_{H^\pm}^{}$. The details are shown
in Ref.~\cite{Ref:AHKOT}.

\begin{figure}[tb]
\includegraphics[height=5.5cm]{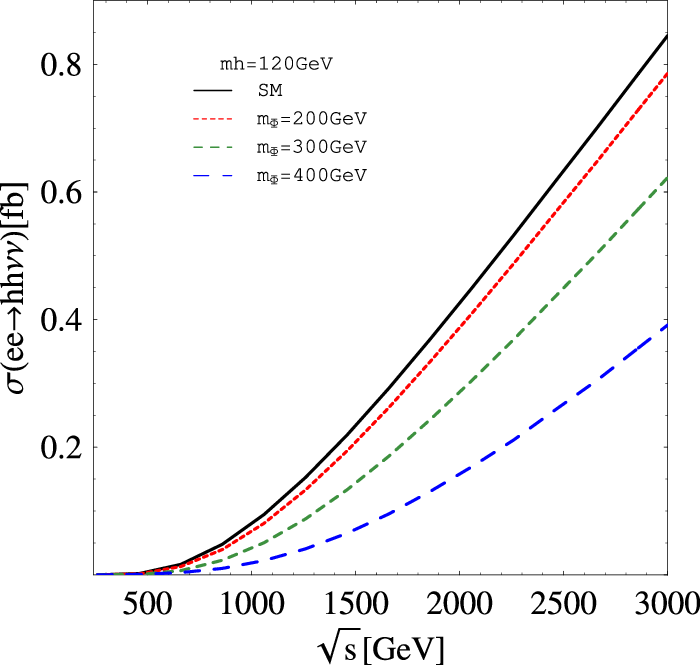}
\includegraphics[height=5.5cm]{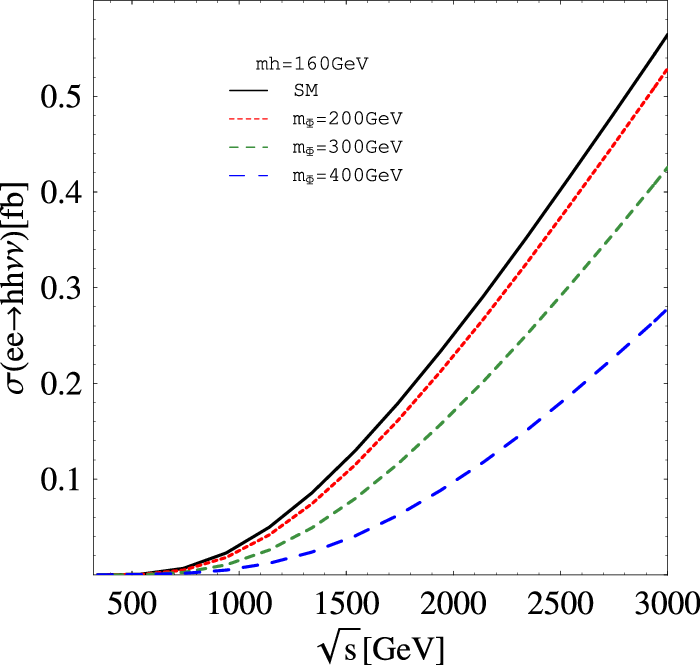}
\caption{The cross sections of $e^+e^-\to hh\nu\bar{\nu}$ process at the ILC as a function of
collision energy $\sqrt{s}$ for $m_h=120$ GeV (left) and $m_h=160$ GeV (right) in the THDM.}
\label{FIG:eehhNNTHDM}
\end{figure}
In FIG.~\ref{FIG:eehhNNTHDM}, the cross sections for $e^+e^-\to hh\nu\bar{\nu}$ with the one-loop
corrected $hhh$ coupling constant due to extra scalars are shown. As we show in
FIG.~\ref{FIG:hhh-thdm}, the $hhh$ coupling constant can deviate from the SM prediction significantly
for $M_{hh}\lesssim 2m_\Phi$, while for $M_{hh}\gtrsim 2m_\Phi$ the deviation becomes small where
$M_{hh}$ varies from $2m_h$ to $\sqrt{s}$. We find that the large corrections in low $M_{hh}$ region
can enhance the cross section by a factor of a few in magnitude. Although the positive one-loop
correction decreases the cross section, this process is still important because the total cross
section can be larger than those in other Higgs pair production processes.

\subsection{Scalar Leptoquarks}
We next consider contributions to cross sections from scalar leptoquarks~\cite{Ref:LQ}.
Unlike the case with the extra Higgs scalar doublet, the scalar leptoquarks are colored fields.
They, therefore, can affect the $gg h$ and the $gg hh$ vertices at the one-loop level, by which
the cross section of $gg\to hh$ in this model can differ from the SM prediction in addition to
the effect of the deviation in the $hhh$ vertex.

We here introduce a complex scalar, leptoquark,
$\phi_\text{\lq}=({\bf \bar 3},{\bf 1})_{1/3} \text{ or } ({\bf \bar 3},{\bf 1})_{4/3}$,
as an example for such theories, where $SU(3)_C$, $SU(2)_L$ and $U(1)_Y$ quantum numbers are shown.
The most general scalar potential can be written as
\begin{align}
V_\text{\lq}(\Phi, \phi_\text{\lq})
&=\lambda\left(\left|\Phi\right|^2-\frac{v^2}2\right)^2
+M_\text{\lq}^2\left|\phi_\text{\lq}\right|^2
+\lambda_\text{\lq}\left|\phi_\text{\lq}\right|^4
+\lambda'\left|\phi_\text{\lq}\right|^2\left|\Phi\right|^2,
\end{align}
where $\Phi$ is the SM Higgs doublet. The mass of leptoquarks is given by
$m^2_{\phi_{LQ}}=M_\text{\lq}^2+\frac{\lambda'v^2}2$.

The searches for the leptoquarks have been performed at the collider experiments at LEP, HERA
and Tevatron. In order to avoid large contributions to lepton flavor violating processes,
the leptoquarks are usually assumed to be coupled only with one fermion generation in their
mass eigenbasis. Under this assumption, the experimental bounds are evaluated as
$m_{\phi_{LQ}}\gtrsim 256$~GeV~\cite{Ref:LQ1}, $316$~GeV~\cite{Ref:LQ2}, and $229$~GeV~\cite{Ref:LQ3}
for the leptoquarks interacting only with the first, second, and third generation, respectively.
There are indirect limits for masses and their Yukawa couplings through the effective
four-fermion interaction~\cite{Ref:PDG}.

In the large mass limit of an $SU(2)_L$ singlet scalar leptoquark, the coupling strength of
the effective $ggh$ vertex is enhanced by the factor of $4/3$ as compared to the SM prediction,
and hence the cross section of the single Higgs boson production $gg\to h$ can be enhanced
approximately by $16/9$. The SM Higgs boson with mass $162$--$166$ GeV has been ruled out
by analyzing $gg\to h\to WW^{(*)}$ process at the Tevatron~\cite{Ref:mh165GeV}. The exclusion band
of the Higgs boson mass from Tevatron results can be translated a wider range as $155$--$185$ GeV.

The one-loop correction to the $hhh$ coupling constant due to the scalar leptoquark can be
calculated analogous to the charged Higgs boson contribution in the THDM~\cite{Ref:KOSY} as
\begin{figure}
\includegraphics[height=5.5cm]{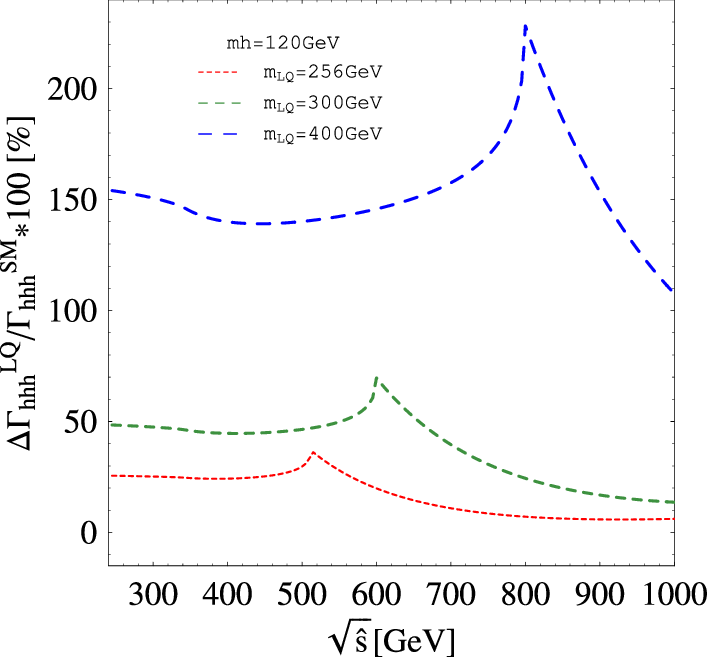}
\includegraphics[height=5.5cm]{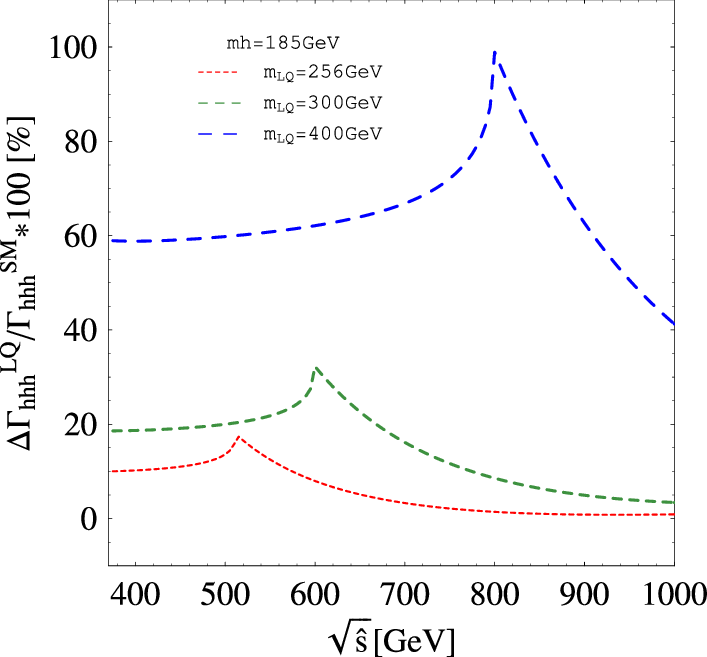}
\caption{The rates for one-loop contributions of an $SU(2)$ singlet scalar leptoquark to
the $hhh$ coupling constant for $m_h=120$ GeV (left) and for $m_h=185$ GeV (right).}
\label{FIG:hhh-LQ}
\end{figure}
\begin{align}
\frac{\Gamma_{hhh}^\text{\lq}}{\Gamma_{hhh}^\text{SM}}
&\simeq 1+\frac{N_cm_{\phi_{LQ}}^4}{6\pi^2v^2m_h^2}
\left(1-\frac{M^2_\text{\lq}}{m_{\phi_{LQ}}^2}\right)^3.
\end{align}
The full expression of the one-loop corrected vertex
$\Gamma_{hhh}^\text{\lq}({\hat s},m_h^2,m_h^2)$ is given in Appendix~A. In FIG.~\ref{FIG:hhh-LQ},
we evaluate the relative size of the one-loop contributions to the $hhh$ coupling constant from
the leptoquarks for $m_h=120$ GeV (left) and for $m_h=185$ GeV (right). Three reference values
are taken for leptoquark masses as $m_{\phi_{LQ}}=256$ GeV (dotted line), $m_{\phi_{LQ}}=300$ GeV
(dashed line), and $m_{\phi_{LQ}}=400$ GeV (long-dashed line). For $m_h=120$ GeV the one-loop correction
can be about $150$\% by a singlet leptoquark with $m_{\phi_{LQ}}=400$ GeV, while for $m_h=185$
GeV it can be about $60$\%. These effects are constructive to the SM value in the nondecoupling
region $M^2_\text{\lq}\simeq 0$, which can be significant for heavy leptoquarks because of the
$m_{\phi_{LQ}}^4$ enhancement. Below the thresholds of $2m_{\phi_{LQ}}$ the quantum effects are
approximately flat for the function of the off-shell Higgs boson energy.

\begin{figure}
\includegraphics[height=5.5cm]{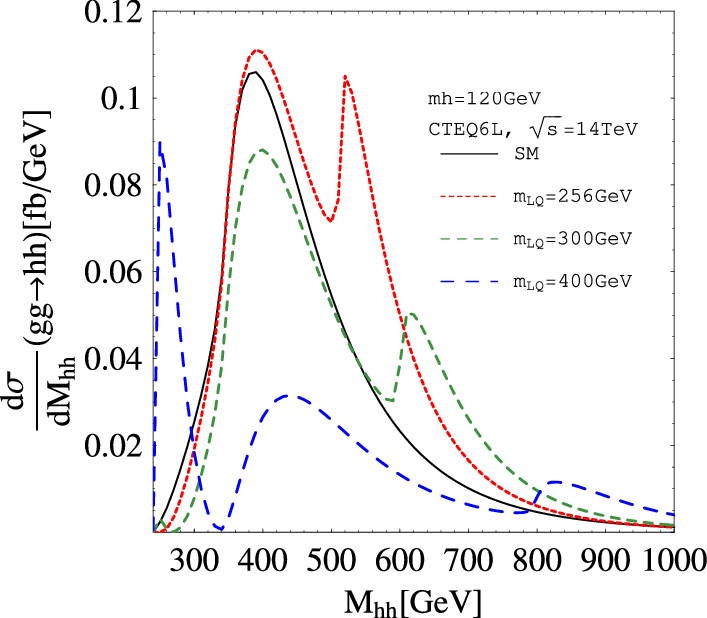}
\includegraphics[height=5.5cm]{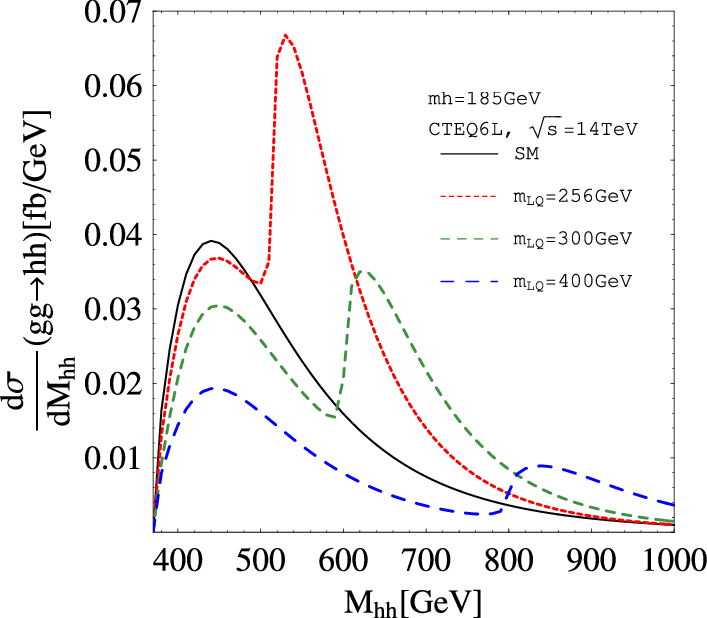}
\caption{The invariant mass distribution of $gg\to hh$ process at the LHC with $\sqrt{s}=14$ TeV
for $m_h=120$ GeV (left) and $m_h=185$ GeV (right) in the leptoquark model.}
\label{FIG:GGhh-inv-LQ}
\end{figure}
In FIG.~\ref{FIG:GGhh-inv-LQ}, we show the invariant mass distribution of the cross section
for $gg\to hh$ process at the LHC. For $m_h=120$ GeV with relatively larger leptoquark masses
$m_\phi\gtrsim 300$ GeV, peaks can be found in the near threshold region of a Higgs boson pair,
which come from the enhancement of the $hhh$ coupling constant as well as that of the $ggh$
coupling constant due to the new colored particles in the triangular diagram: see
FIG.~\ref{FIG:diag-gghh} for the reference. The threshold enhancement of the on-shell leptoquark
pair production can also be seen around $M_{hh}\simeq 2m_\phi$.

\begin{figure}
\includegraphics[height=5.5cm]{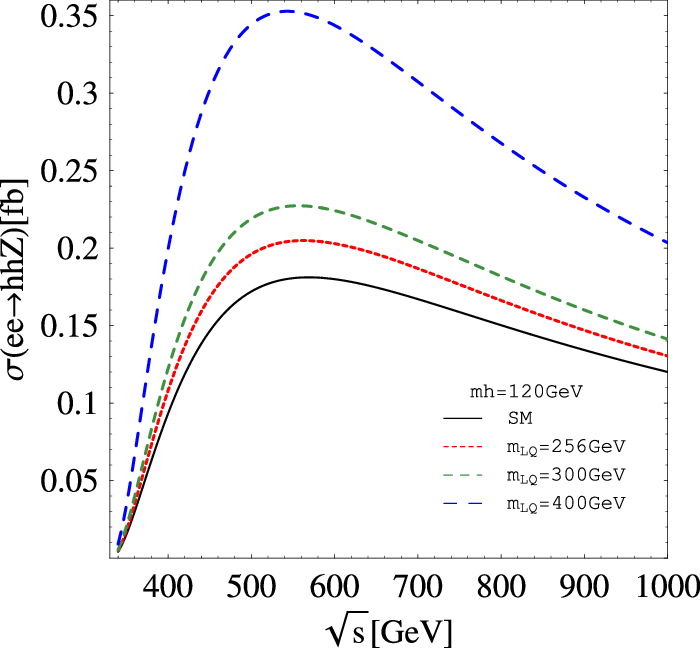}
\includegraphics[height=5.5cm]{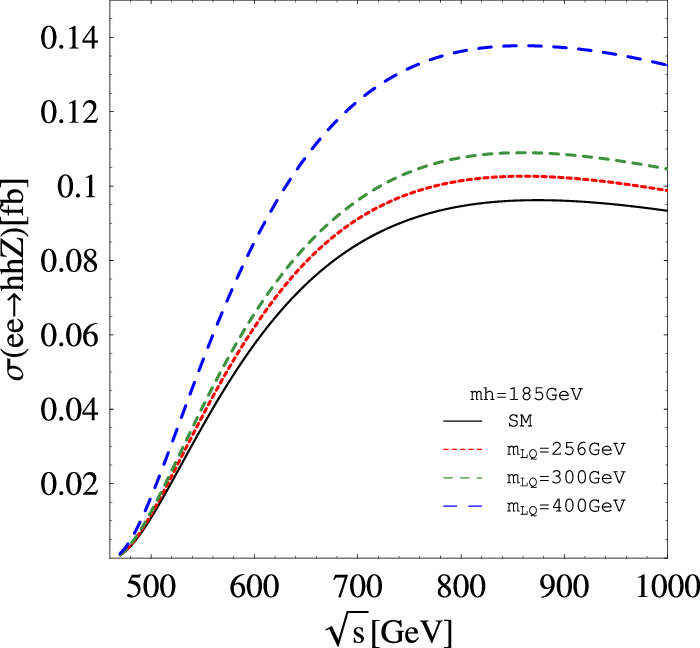}
\caption{The cross section of $e^+e^-\to hhZ$ process as a function of $\sqrt{s}$
for $m_h=120$ GeV (left) and $m_h=185$ GeV (right) in leptoquark models.}
\label{FIG:eehhZ-LQ}
\end{figure}
In FIG.~\ref{FIG:eehhZ-LQ}, the cross sections for $e^+e^-\to hhZ$ are shown in the case with
$m_h=120$ GeV as a function of the collision energy $\sqrt{s}$ for various $m_{\phi_{LQ}}$.
The effects of the scalar leptoquarks only appear in the $hhh$ coupling constant.

\begin{figure}
\includegraphics[height=5.5cm]{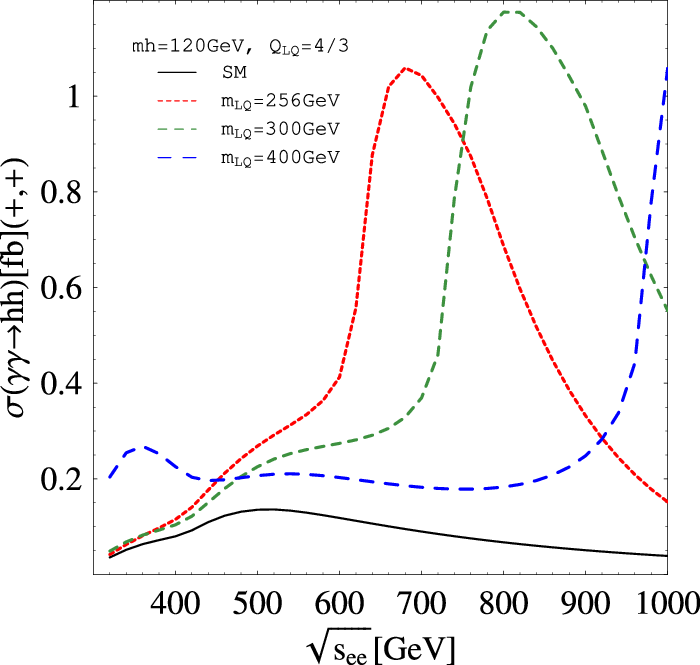}
\includegraphics[height=5.5cm]{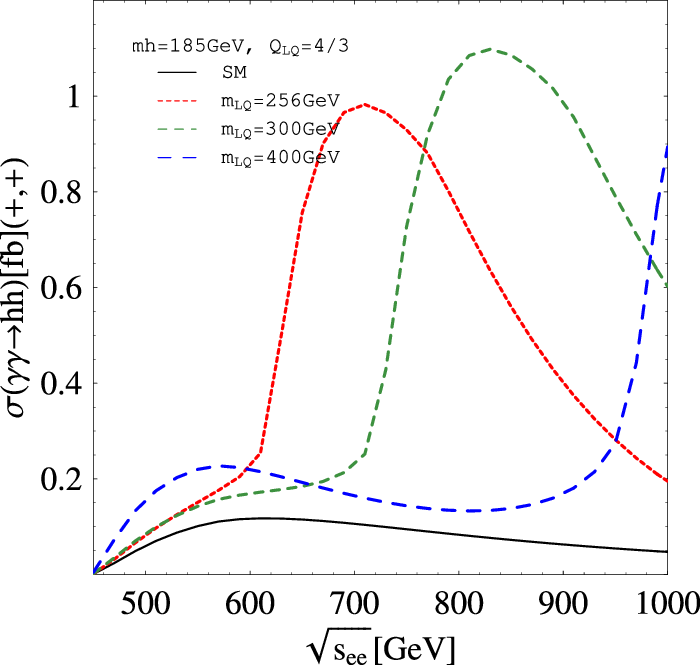}
\includegraphics[height=5.5cm]{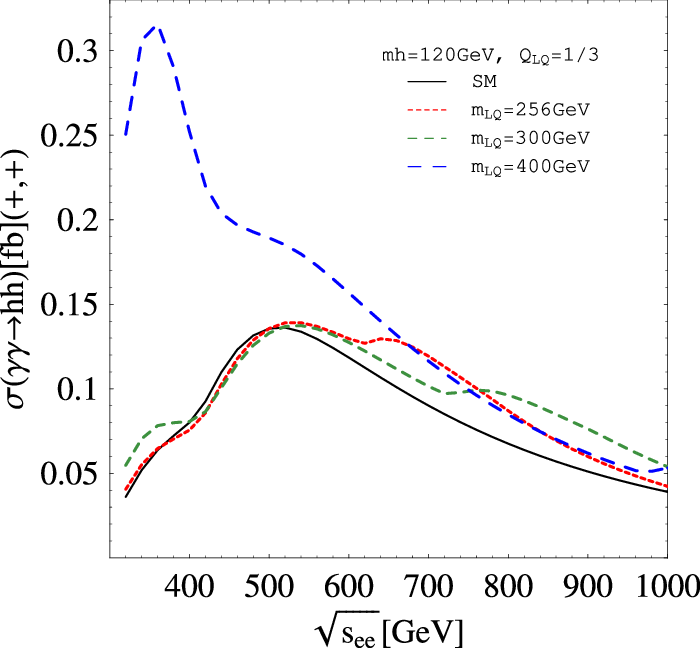}
\includegraphics[height=5.5cm]{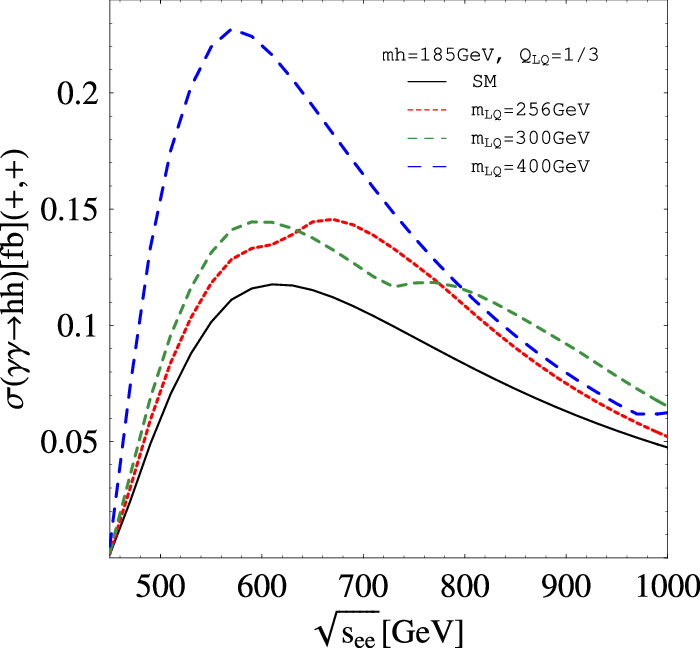}
\caption{The cross section of $\gamma(+)\gamma(+)\to hh$ process at the photon collider option
at the ILC as a function of the $e^-e^-$ collision energy for $m_h=120$ GeV (left column) and
$m_h=185$ GeV (right column). The electric charges for scalar leptoquarks are taken as $Q=4/3$ (top)
and $Q=1/3$ (bottom), respectively.}
\label{FIG:GamGamhh-LQ}
\end{figure}
In FIG.~\ref{FIG:GamGamhh-LQ}, we show the cross section of the Higgs pair production process
at a PLC. The effects of the scalar leptoquarks depend on not only their masses but also their
electric charges. The first peaks around the threshold region come from the modifications of
the $hhh$ coupling constant and the effective $\gamma\gamma h$ vertex due to the scalar leptoquarks.
The former effect does not depend on the quantum number, while the latter does. The threshold
enhancements of the box diagram can be found after $\sqrt{s_{ee}}\sim 2m_\phi$ which are also
dependent on the electric charges.

\begin{figure}[tb]
\includegraphics[height=5.5cm]{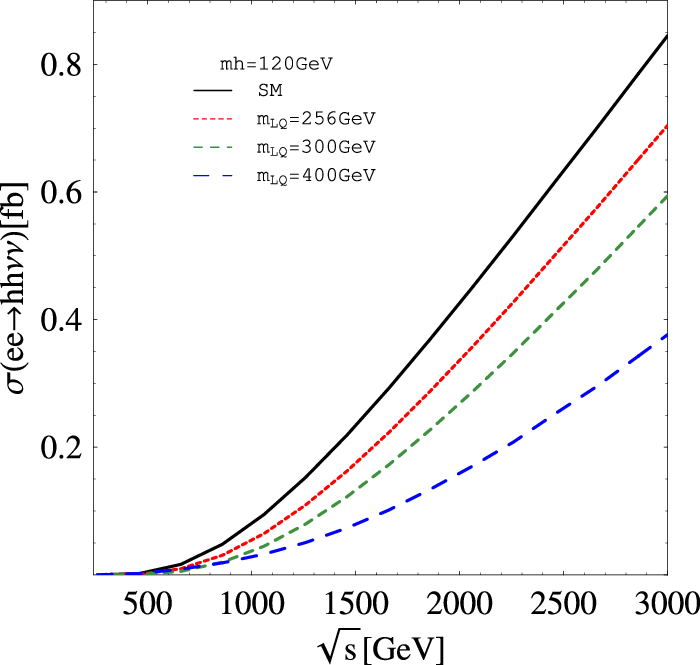}
\includegraphics[height=5.5cm]{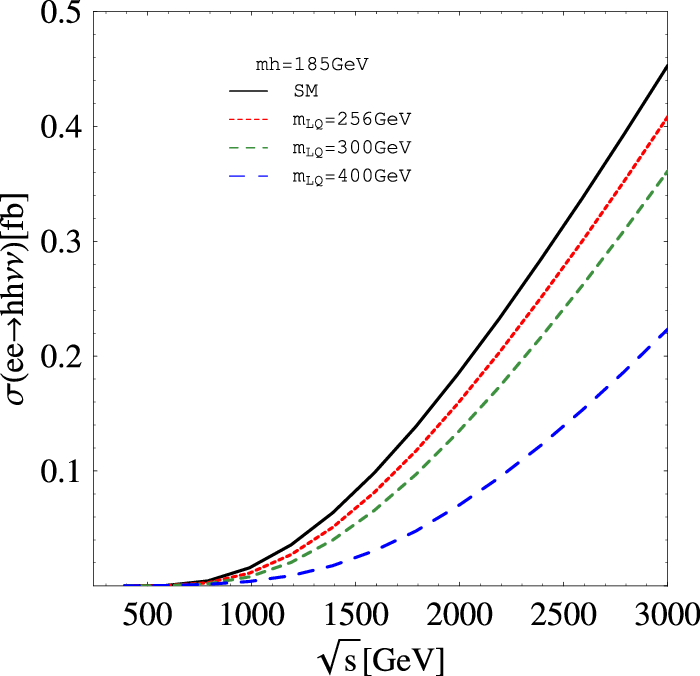}
\caption{The cross sections of $e^+e^-\to hh\nu\bar{\nu}$ process at the ILC as a function of
collision energy $\sqrt{s}$ for $m_h=120$ GeV (left) and $m_h=185$ GeV (right) in leptoquark models.}
\label{FIG:eehhNNLQ}
\end{figure}
In FIG.~\ref{FIG:eehhNNLQ}, the cross section for $e^+e^-\to hh\nu\bar{\nu}$ is shown for $m_h=120$
GeV (left) and $m_h=185$ GeV (right) with the $hhh$ coupling corrections of leptoquarks.
The cross section becomes smaller as in the THDM because of the negative interference.

\subsection{Chiral fourth generation}
One of the fundamental questions in the SM is the number of generation (family) of quarks
and leptons. There is no theoretical reason to restrict the fermion families to be three.
The electroweak precision data also do not exclude completely existence of the sequential
fourth generation. The fourth generation quarks are colored particles so that they, similarly
to the case of scalar leptoquarks, affect the $ggh$ and the $gghh$ vertices. In addition,
the chiral fermion has the nondecoupling property; i.e., the mass is completely proportional
to the VEV, so that the one-loop correction to the $hhh$ coupling constant can be very large
in magnitude.

We here introduce a sequential set of fermions, i.e., $Q'=({t'}_L^{},{b'}_L^{})^T,
L'=({\ell'}_L^{},{\nu'}_L^{})^T, {t'}_R^{}, {b'}_R^{}, {\ell'}_R^{}, {\nu'}_R^{}$,
as the chiral fourth generation model (Ch4). Neutrinos are assumed to be Dirac particle
whose masses are generated by Yukawa interaction.
From LEP results, a lower bound on the extra charged lepton $(\ell')$ mass is $100.8$ GeV,
while for heavy neutral lepton $(\nu')$ with Dirac nature to be $90.3$ GeV. The fourth
generation up-type quark $(t')$ is rather stringently constrained, $m_{t'}\gtrsim 256$ GeV,
by Tevatron. This bound is independent of the Cabibbo-Kobayashi-Maskawa (CKM) elements
between the first three SM fermions and the fourth generation. A similar limit for the heavy
down-type quark $(b')$ is obtained as $m_{b'}\gtrsim 128 (268)$ GeV by using the charged
(neutral) current decay modes~\cite{Ref:PDG}. Further stronger bounds can be found in
Ref.~\cite{Ref:4gmass} by assuming additional assumptions.

The contributions from the fourth generation fermions to the oblique electroweak parameters can
be significant. In the limit of heavy fermions, the ${\widehat S}$-parameter is calculated as
\begin{align}
{\widehat S}=\frac2{3\pi}-\frac1{6\pi}\left(\ln\frac{m_{t'}^2}{m_{b'}^2}
-\ln\frac{m_{\nu'}^2}{m_{\ell'}^2}\right).
\end{align}
It is noted that a complete set of the four generation fermions with degenerate masses
is excluded at $6\sigma$ level, if we take into account the constraint only from
the ${\widehat S}$-parameter~\cite{Ref:PDG,Erler:2010sk}. It can be relaxed by requiring
the preferable mass hierarchy of the fourth generation fermions i.e., $m_{t'}\gtrsim m_{b'}$
and $m_{\ell'}\gtrsim m_{\nu'}$~\cite{Ref:Kribs4G}. Furthermore, the breaking of isospin symmetry
in the fourth generation doublet gives substantial positive contribution to
the ${\widehat T}$-parameter, which pulls back the model to the allowed region in
$({\widehat S},{\widehat T})$ plane. For $m_h=117$ GeV, the  $95\%$ CL upper bounds are given as
$\widehat{S}\le0.16$ and $\widehat{T}\le0.21$ with a strong correlation~\cite{Ref:PDG,Erler:2010sk}.
In Table~\ref{Tab:4G-ST}, the contributions from the fourth generation fermions to oblique
parameters are listed. In order to reduce the number of parameters, $m_{t'}=m_{\ell'}$ and
$m_{b'}=m_{\nu'}$ are taken. We can see that the appropriate mass difference
($m_{t'}-m_{b'}=50$--$55$ GeV) between the fourth generation fermions fits the constraint from
the electroweak precision data.

\begin{table}[t]
\begin{tabular}{|c||c|c|c|}
\hline
$({\widehat S},{\widehat T})$& $m_{t'}-m_{b'}=50$ GeV& $m_{t'}-m_{b'}=55$ GeV& $m_{t'}-m_{b'}=60$ GeV \\
\hline\hline
$m_{t'}= 256$ GeV & $(0.18,0.18)$ & $(0.18,0.22)$ & $(0.17,0.26)$ \\ \hline
$m_{t'}= 300$ GeV & $(0.19,0.18)$ & $(0.18,0.22)$ & $(0.18,0.26)$ \\ \hline
$m_{t'}= 400$ GeV & $(0.19,0.18)$ & $(0.19,0.22)$ & $(0.19,0.26)$ \\ \hline
\end{tabular}
\caption{
The contributions from the chiral fourth generation fermions to ${\widehat S}$ and ${\widehat T}$
are shown. The mass degeneracies $m_{t'}=m_{\ell'}$ and $m_{b'}=m_{\nu'}$ are assumed.}
\label{Tab:4G-ST}
\end{table}

From a theoretical point of view, there are a few constraints such as triviality bounds and
vacuum stability bounds on the coupling constants~\cite{Ref:Hachimoto4G}. To avoid the Landau pole
for the fourth generation Yukawa coupling constant, the mass should be lighter than about $577$ GeV
when the cutoff scale of theory is taken to be $2$ TeV. Instability of the vacuum gives more serious
bound on the Higgs boson mass. It may be evaded by introduction of the extra Higgs
doublet~\cite{Ref:Hachimoto4G} or some other new physics dynamics, and hence we here keep the mass
of the (SM-like) Higgs boson to be the electroweak scale.

The additional heavy colored particles enhance the effective $ggh$ coupling approximately by a factor
of $3$ which leads to enhancement of the cross section of $gg\to h$ by a factor of $9$ at hadron
colliders. Tevatron bounds on the SM Higgs boson mass~\cite{Ref:mh165GeV} can be translated into
a wider exclusion band as $125$--$200$ GeV for the Higgs boson mass in the chiral fourth generation
model~\cite{Ref:planck2010}.

The large one-loop effect on the $hhh$ coupling constant in the SM is generalized straightforwardly
to the chiral fourth generation model:
\begin{align}
\frac{\Gamma_{hhh}^\text{Ch4}}{\Gamma_{hhh}^\text{SM}}
&\simeq 1-\sum_{f'=t',b',\ell',\nu'}\frac{N_cm_{f'}^4}{3\pi^2v^2m_h^2}.
\end{align}
The explicit formula of $\Gamma_{hhh}^\text{Ch4}$ including energy dependence is given in Appendix~A.
Since $m_{f'}^4$ enhancements come from extra heavy fermions, we would expect large quantum corrections
to the $hhh$ coupling constant. We note that these fermion loop contributions are always negative
to the SM prediction.
\begin{figure}
\includegraphics[height=5.5cm]{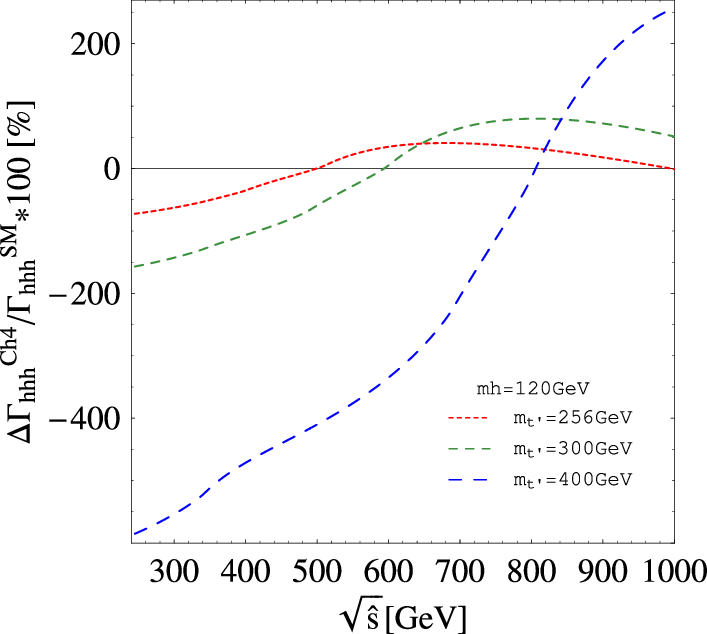}
\includegraphics[height=5.5cm]{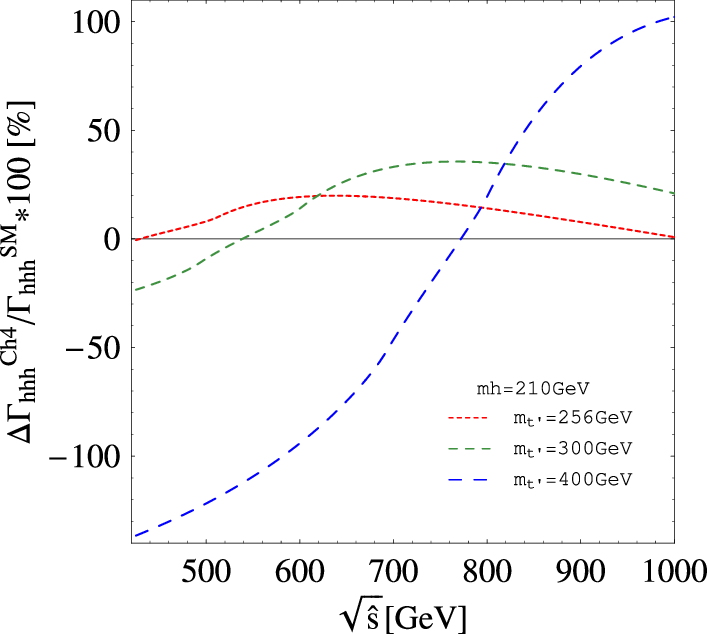}
\caption{The rates for one-loop contributions of the chiral fourth generation fermions to
the $hhh$ coupling constant for $m_h=120$ GeV (left) and for $m_h=210$ GeV (right). The dotted,
dashed, long-dashed curved lines indicate masses of heavy fermions as $m_{t'}=256, 300, 400$ GeV,
respectively, with the appropriate mass difference $m_{t'}-m_{b'}=55$ GeV.}
\label{FIG:hhh-ch4}
\end{figure}
In FIG.~\ref{FIG:hhh-ch4}, effects of the chiral fourth generation fermions on the $hhh$ coupling
constant are shown for $m_h=120$ GeV (left) and for $m_h=210$ GeV (right). Hereafter, the mass
differences are fixed to be $m_{t'}-m_{b'}=55$ GeV with $m_{t'}=m_{\ell'}$ and $m_{b'}=m_{\nu'}$.
The masses of the fourth generation up-type quark are taken as three representative values,
$m_{t'}=256$ GeV (dotted line), $300$ GeV (dashed line), and $400$ GeV (long-dashed line). The $hhh$ coupling
constant is changed significantly depending on the energy of the off-shell Higgs boson.
In the low energy limit, a huge quantum correction to the $hhh$ coupling constant can be more than
$100$\%, which can easily overwhelm the SM contribution and change the sign of the total amplitude.
We again note that from the vacuum stability condition it is highly disfavored to introduce too
heavy fourth generation fermion unless the Higgs sector is extended.

\begin{figure}
\includegraphics[height=5.5cm]{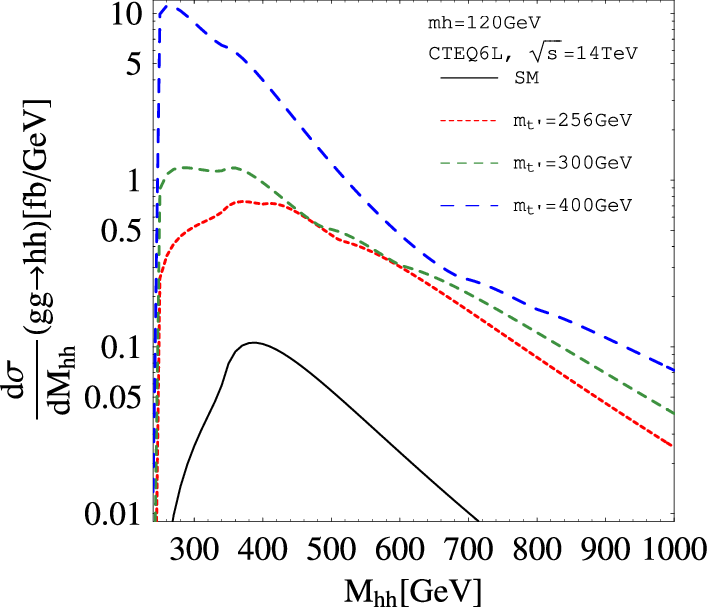}
\includegraphics[height=5.5cm]{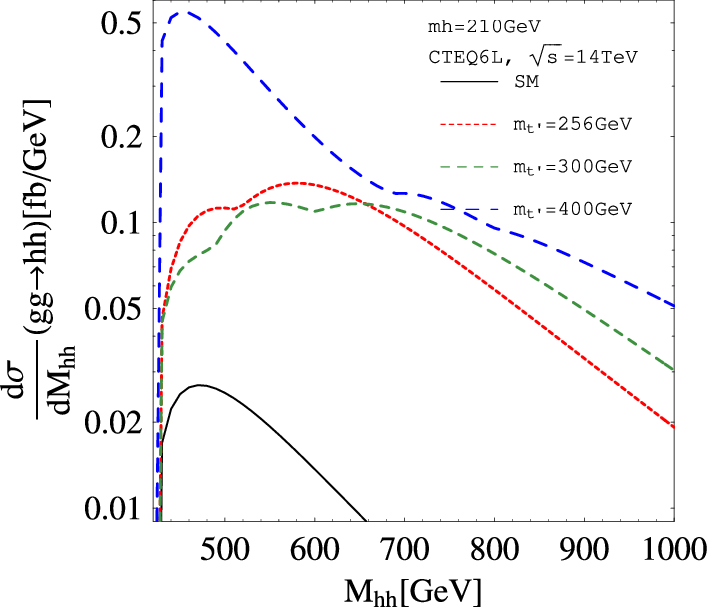}
\caption{The invariant mass distribution of $gg\to hh$ process at the LHC with $\sqrt{s}=14$ TeV
for $m_h=120$ GeV (left) and $m_h=210$ GeV (right) in the chiral fourth generation model. Three
representative values of $t'$ mass are chosen as $256$ GeV (dotted line), $300$ GeV (dashed line), and $400$
GeV (long-dashed line). The SM prediction is also shown by a solid curved line for comparison.}
\label{FIG:GGhh-inv-ch4}
\end{figure}
In FIG.~\ref{FIG:GGhh-inv-ch4}, we show the invariant mass distribution of the differential cross
section of $gg\to hh$ in the SM with a complete set of fourth generation for $m_h=120$ GeV (left)
and for $m_h=210$ GeV (right). The production cross sections of the chiral fourth generation model
can be $10$--$100$ times larger than that of the SM. The first peaks come not only from the large
one-loop correction to the $hhh$ vertex but also from the enhancement of the $ggh$ vertex due to
the fourth generation quarks. The threshold enhancements of on-shell fermion-pair production can also
be seen around $M_{hh}\simeq 2m_t$ and $M_{hh}\simeq 2m_{f'}$ which are smeared by the distribution
of high energy gluons.

\begin{figure}
\includegraphics[height=5.5cm]{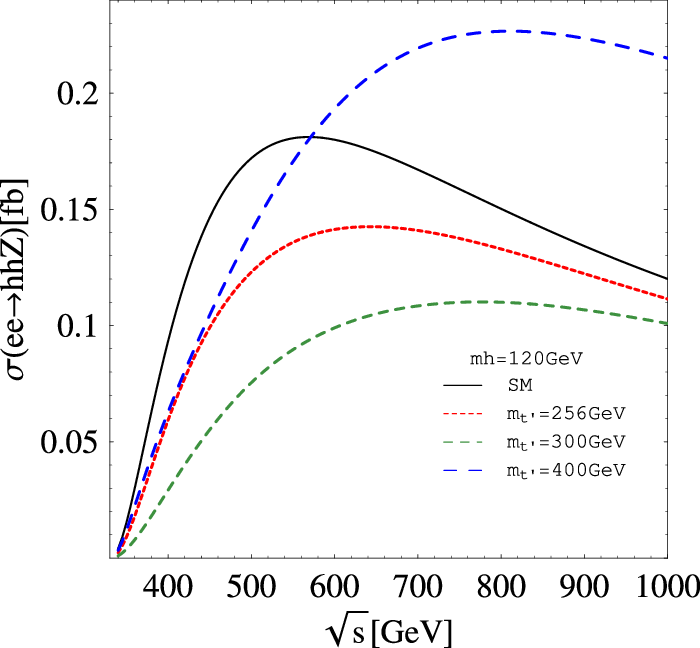}
\includegraphics[height=5.5cm]{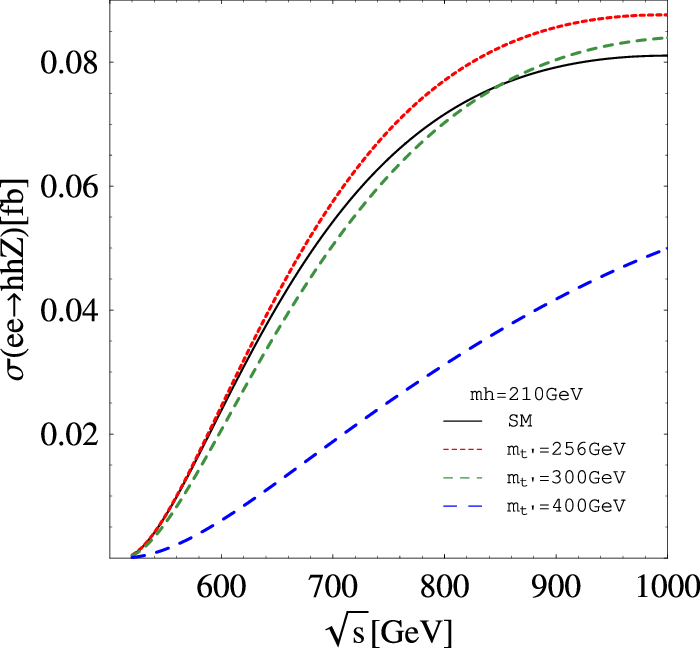}
\caption{The cross section of $e^+e^-\to hhZ$ process as a function of $\sqrt{s}$ for $m_h=120$ GeV
(left) and $m_h=210$ GeV (right) in the chiral fourth generation model. }
\label{FIG:eehhZ-ch4}
\end{figure}
In FIG.~\ref{FIG:eehhZ-ch4}, we show the cross section of the double-Higgs-strahlung process
$e^+e^-\to hhZ$ as a function of the collision energy. The cross section can be reduced by
the suppression of the $hhh$ coupling constant due to the fourth generation fermions. Unlike
the case of $\Delta\kappa$ approximation the deviation of the cross section from the SM depends
on the collision energy, because the quantum corrections to the $hhh$ coupling constant are
the function of the energy for the off-shell Higgs boson.

\begin{figure}
\includegraphics[height=5.5cm]{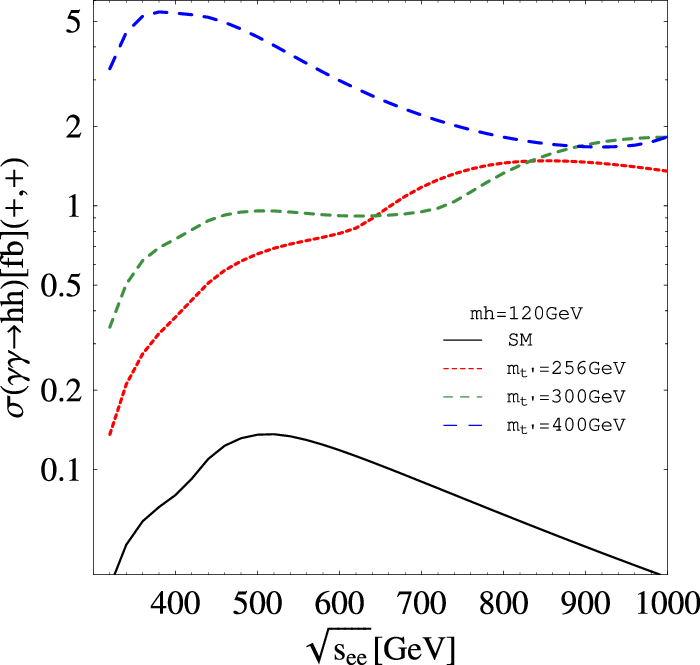}
\includegraphics[height=5.5cm]{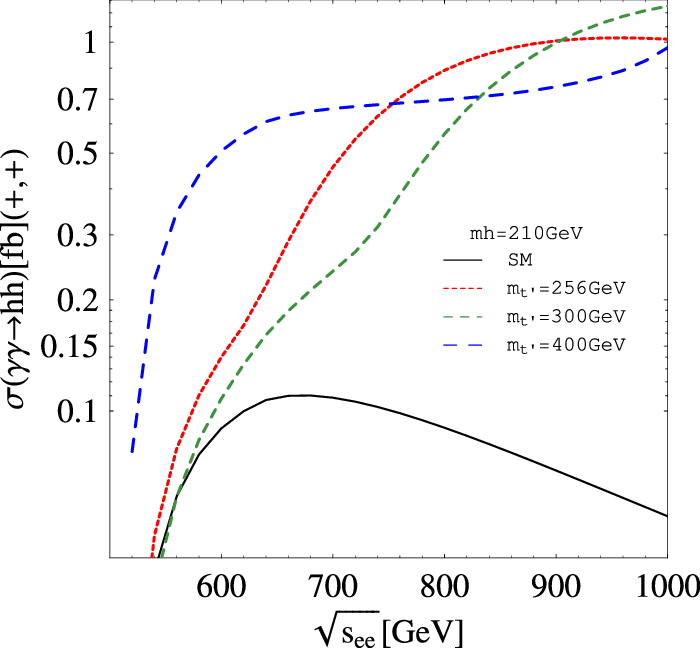}
\caption{The cross section of $\gamma(+)\gamma(+)\to hh$ process at the photon collider option
at the ILC as a function of the $e^-e^-$ collision energy for $m_h=120$ GeV (left) and $m_h=210$
GeV (right) in the chiral fourth generation model.}
\label{FIG:GamGamhh-ch4}
\end{figure}
In FIG.~\ref{FIG:GamGamhh-ch4}, the cross section of the Higgs pair production at a photon
collider is given as a function of $\sqrt{s_{ee}}$ with $m_h=120$ GeV (left) and $m_h=210$
GeV (right). All the fourth generation fermions contribute to the both triangular and the box
diagrams (see FIG.~\ref{FIG:diag-gamgamhh}) due to the large Yukawa coupling constant, which
can enhance the cross section significantly by a factor of $10$ for the wide range of $\sqrt{s_{ee}}$.
The threshold effects of the on-shell heavy fermions can be found soon above the thresholds
$\sqrt{s_{ee}}\sim 2m_t$ and $\sqrt{s_{ee}}\sim 2m_{f'}$ because the photon luminosity has
a peak around $0.8\sqrt{s_{ee}}$. At the PLC, we can have larger cross sections even for
relatively heavy Higgs bosons.

\begin{figure}[tb]
\includegraphics[height=5.5cm]{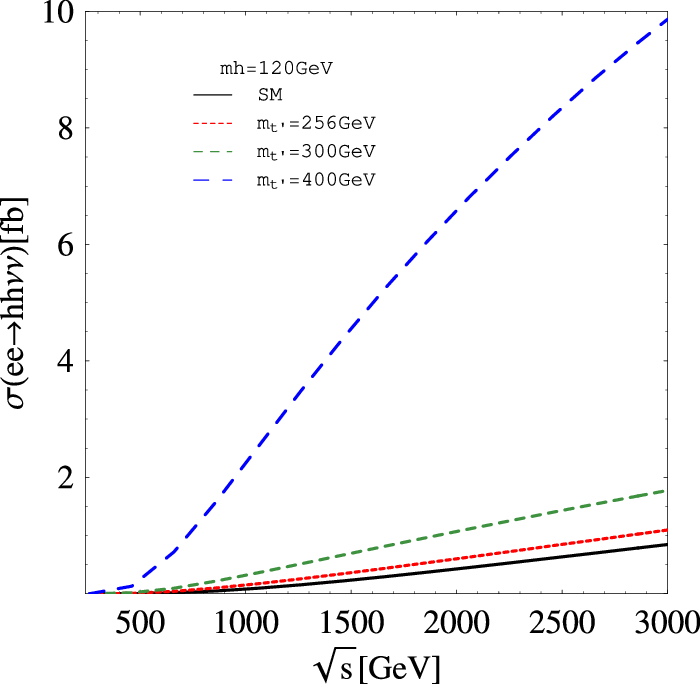}
\includegraphics[height=5.5cm]{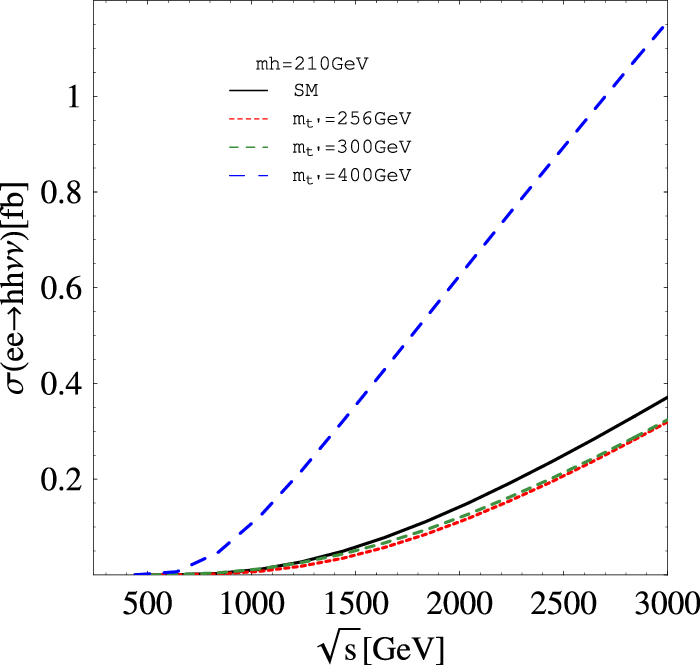}
\caption{The cross sections of $e^+e^-\to hh\nu\bar{\nu}$ process at the ILC as a function
of collision energy $\sqrt{s}$ for $m_h=120$ GeV (left) and $m_h=210$ GeV (right) in the
chiral fourth generation model.}
\label{FIG:eehhNNCh4}
\end{figure}
In FIG.~\ref{FIG:eehhNNCh4}, we show the cross section for $e^+e^-\to hh\nu\bar{\nu}$ with
the one-loop corrected $hhh$ coupling constant. For $m_h=120$ GeV (left), the production rate
becomes significantly large compared to the SM rate. This enhancement mainly comes from
the large quantum corrections to the $hhh$ coupling constant in the smaller $M_{hh}$ region.
For $m_h=210$ GeV, corrections to the $hhh$ coupling constant are relatively small. However,
for larger $M_{hh}$ values the one-loop correction
$(\Delta\Gamma^\text{Ch4}_{hhh}/\Gamma^\text{SM}_{hhh})$ goes back to negative, and its effect
rapidly becomes important for $M_{hh}\gtrsim 1500$ GeV, which makes the cross section larger.

\subsection{Vectorlike quarks}
Various types of vectorlike fermions have also been discussed in the literature. They can appear
in extra-dimension models with bulk fermions~\cite{Ref:ExD}, in little Higgs models~\cite{Ref:LH}
and in the top seesaw model~\cite{Ref:topseesaw}. As a representative case of these models,
we adopt a pair of vectorlike up-type quarks, $T_{0L}$ and $T_{0R}$, which transform as
$({\bf 3}, {\bf 1})_{2/3}$ under the gauge symmetry.

The Lagrangian relevant to the mass of the SM top-quark and the vectorlike up-type quark can be
written as
\begin{align}
{\mathcal L}^\text{mass}
&= -y_t\,\overline{Q}_0\,t_{0R}^{}\,\widetilde{\Phi}
-Y_{T}\,\overline{Q}_0\,T_{0R}\,\widetilde{\Phi}-M_{T}\,\overline{T}_{0L}^{}T_{0R}
+\text{H.c.},
\end{align}
where we have dropped the terms proportional to $\overline{T}_{0L}^{}t_{0R}^{}$ which are absorbed
by redefinitions of $t_{0R}^{}$ and $T_{0R}$ without loss of generality. Since the $t_0$--$T_0$
mixing term is allowed by the symmetry, $t_0$ and $T_0$ are no longer mass eigenstates.
The mass eigenstates $t$ and $T$ are determined by diagonalization of the mass matrix,
\begin{align}
{\widehat M}_t
=\begin{pmatrix}\frac{y_t\,v}{\sqrt2}&\frac{Y_T\,v}{\sqrt2}\\
0&M_T\end{pmatrix}
&
= U_L^\dag\begin{pmatrix}m_t&\\&m_T\end{pmatrix}U_R,
\end{align}
where
\begin{align}
\begin{pmatrix}t_X^{}\\T_X\end{pmatrix}
=U_{X}\begin{pmatrix}t_{0X}^{}\\T_{0X}\end{pmatrix}
=\begin{pmatrix}c_{X}^{}&-s_{X}^{}\\s_{X}^{}&c_{X}^{}\end{pmatrix}
\begin{pmatrix}t_{0X}^{}\\T_{0X}\end{pmatrix},
\quad \text{ where } X=L,R.
\end{align}

The direct search for the vectorlike quarks has been performed~\cite{Ref:PDG}. Their main
production modes at hadron colliders are $gg\to\overline{T}T$, so that the lower bound of
the up-type vectorlike fermion is basically the same as the fourth generation up-type quark,
$m_T^{}\gtrsim 256$ GeV.

The vectorlike quarks are also severely constrained by electroweak precision data~\cite{Ref:Vec}.
For ${\widehat U}=0$, the experimental values for oblique parameters are ${\widehat S}=-0.04\pm0.09$
and ${\widehat T}=0.02\pm0.09$ where $m_h=117$ GeV is assumed. The contributions to the ${\widehat S}$
parameter due to the vectorlike top-quark $(T)$ are less than $0.004$ for $M_T\gtrsim 1000$ GeV,
which can be neglected. The $t_0-T_0$ Yukawa coupling constant is set to be $Y_T=1$ throughout this
analysis. On the other hand, the ${\widehat T}$ parameter is rather sensitive to the model parameters.
The one (two) sigma bound on the lowest value of $M_T$ is $1100$ (1700) GeV for $m_h=120$ GeV.
For $m_h=160$ GeV, these constraints are slightly milder, $M_T\gtrsim 1100 (1500)$ GeV at
$1\sigma (2\sigma)$ confidence level.

The one-loop correction to the $hhh$ coupling constant due to the vectorlike top-quark is evaluated as
\begin{figure}
\includegraphics[width=5.5cm]{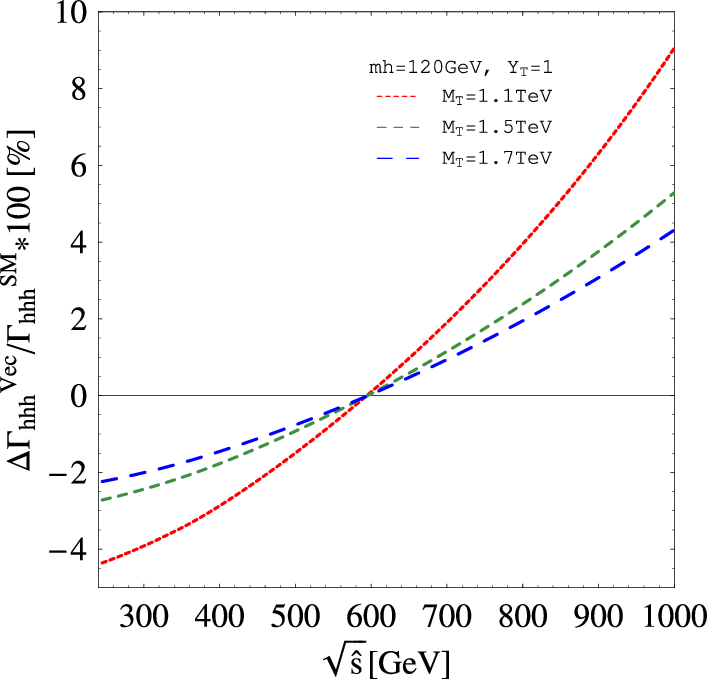}
\includegraphics[width=5.5cm]{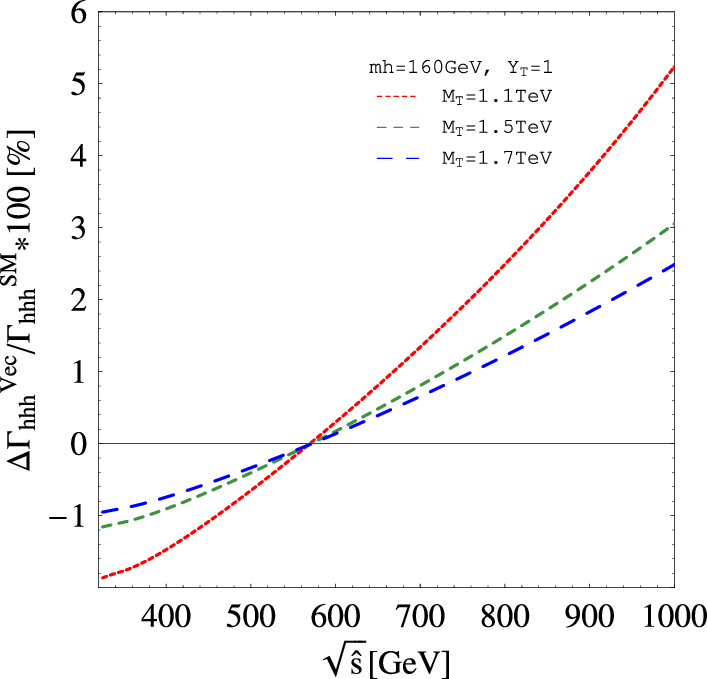}
\caption{The rates for one-loop contributions from the vectorlike top-quark $T$ to the $hhh$ coupling
constant for $m_h=120$ GeV (left) and for $m_h=160$ GeV (right). The $t_0$--$T_0$ Yukawa coupling is
taken to be $Y_T=1$, and the gauge invariant mass parameter $M_T$ is chosen as $1100$ GeV (dotted line),
$1500$ GeV (dashed line) and $1700$ GeV (long-dashed line), respectively.}
\label{FIG:hhh-vf}
\end{figure}
\begin{align}
\frac{\Gamma_{hhh}^\text{Vec}}{\Gamma_{hhh}^\text{SM}}
&\simeq 1
-\frac{N_cm_T^4}{3\pi^2v^2m_h^2(1-m_t^2/m_T^2)}
\left(1-\frac{y_t^\text{eff}}{\sqrt2}\frac{v}{m_t}\right)\Biggl\{ 
%
-\frac{3m_t^4}{m_T^4}\left[3-\frac{m_t^2}{m_T^2}-\frac2{1-m_t^2/m_T^2}\ln\frac{m_T^2}{m_t^2}\right]\nonumber\\
&\qquad+\frac{3m_t^2}{m_T^2}
\left[2+5\frac{m_t^2}{m_T^2}-\frac{m_t^4}{m_T^4}-\frac{6m_t^2/m_T^2}{1-m_t^2/m_T^2}\ln\frac{m_T^2}{m_t^2}\right]
\left(1-\frac{y_t^\text{eff}}{\sqrt2}\frac{v}{m_t}\right)\nonumber\\
&\qquad+\left[\left(1+\frac{m_t^2}{m_T^2}\right)\left(1-8\frac{m_t^2}{m_T^2}+\frac{m_t^4}{m_T^4}\right)
+\frac{12m_t^4/m_T^4}{1-m_t^2/m_T^2}\ln\frac{m_T^2}{m_t^2}\right]
\left(1-\frac{y_t^\text{eff}}{\sqrt2}\frac{v}{m_t}\right)^2
\Biggr\},
\end{align}
where $y_t^\text{eff}=c_L^{}(c_R^{}y_t-s_R^{}Y_T)$. Although there is a $m_T^4$ enhancement factor,
the correction to the $hhh$ coupling constant can not be large. This is because a factor
$\left(1-y_t^\text{eff}v/(\sqrt2m_t)\right)$ is approximately expressed as
$\left(Y_T v/(\sqrt2 m_T)\right)^2$ for large $m_T^{}$, so that the correction to the $hhh$ coupling
constant decouples as $1/m_T^2$.
In FIG.~\ref{FIG:hhh-vf}, the effects on the $hhh$ coupling constant due to the vectorlike top-quark
are shown. The stringent experimental bounds from the electroweak precision data impose that mass of
$T$ particle is heavy, $m_T\gtrsim 1100$ GeV. Therefore, there can be no significant nondecoupling
effect on the $hhh$ coupling constant.

\begin{figure}
\includegraphics[width=5.5cm]{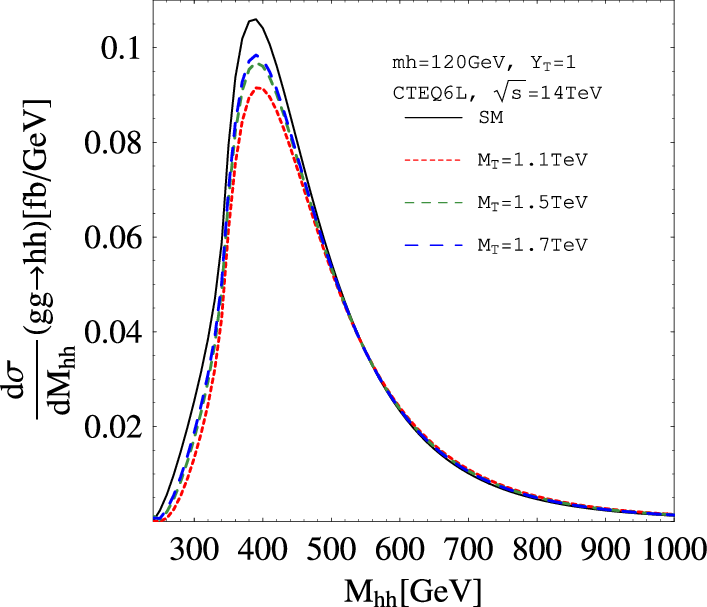}
\includegraphics[width=5.5cm]{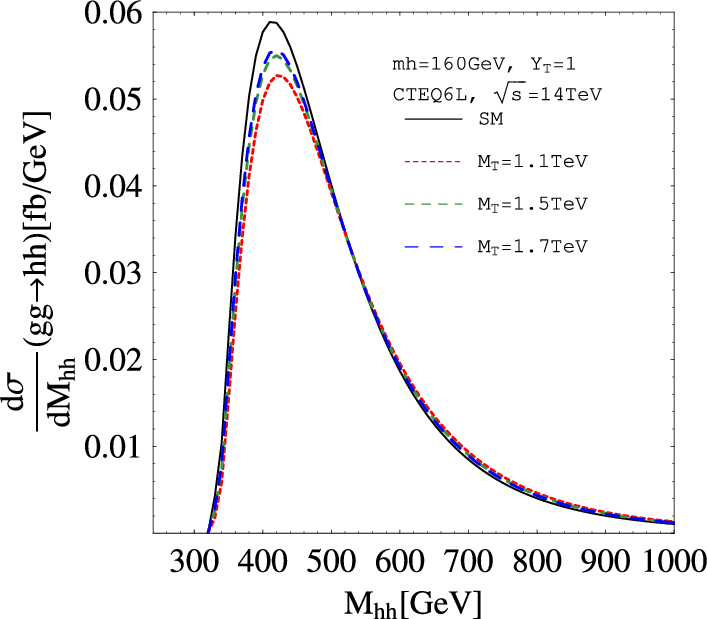}
\caption{The invariant mass distribution of $gg\to hh$ process at the LHC with $\sqrt{s}=14$ TeV
for $m_h=120$ GeV (left) and $m_h=160$ GeV (right) in the SM with the vectorlike top-quark.}
\label{FIG:GGhh-inv-vf}
\end{figure}
In FIG.~\ref{FIG:GGhh-inv-vf}, we show the invariant mass distribution for the cross section of
$gg\to hh$ at the LHC with $\sqrt{s}=14$ TeV in the model with the vectorlike top-quark.
The vectorlike top-quarks give new contribution to the both triangular and box diagrams. However,
the deviations of the cross section from the SM value are at most $5\%$, which can not be large
because of their decoupling nature.
%
\begin{figure}
\includegraphics[width=5.5cm]{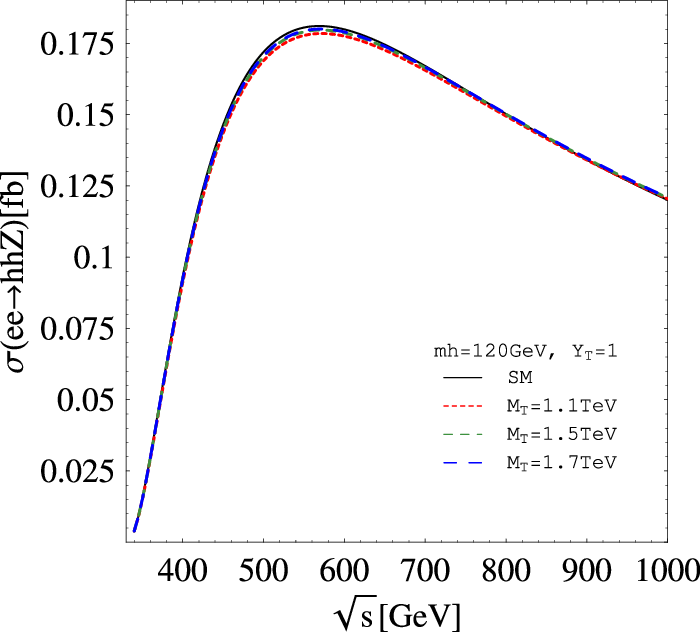}
\includegraphics[width=5.5cm]{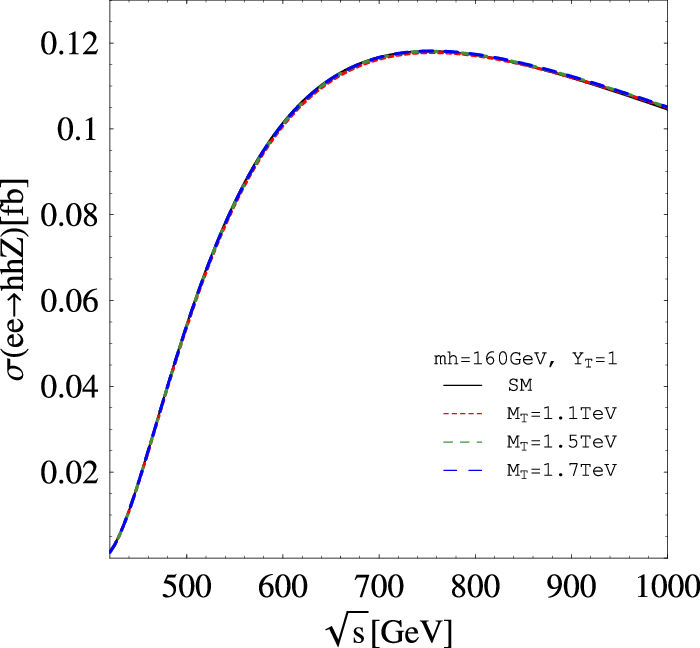}
\caption{The cross section of $e^+e^-\to hhZ$ process as a function of $\sqrt{s}$ for $m_h=120$ GeV
(left) and $m_h=160$ GeV (right) in the SM with the vectorlike top-quark.}
\label{FIG:eehhZ-vf}
\end{figure}
In FIG.~\ref{FIG:eehhZ-vf}, the cross sections for the double-Higgs-strahlung process are shown in
the model with the vectorlike top-quark. The effects of the vectorlike fermions only appear in
the $hhh$ coupling constant. The impact of the vectorlike top-quarks is quite small also in this process.
%
\begin{figure}
\includegraphics[width=5.5cm]{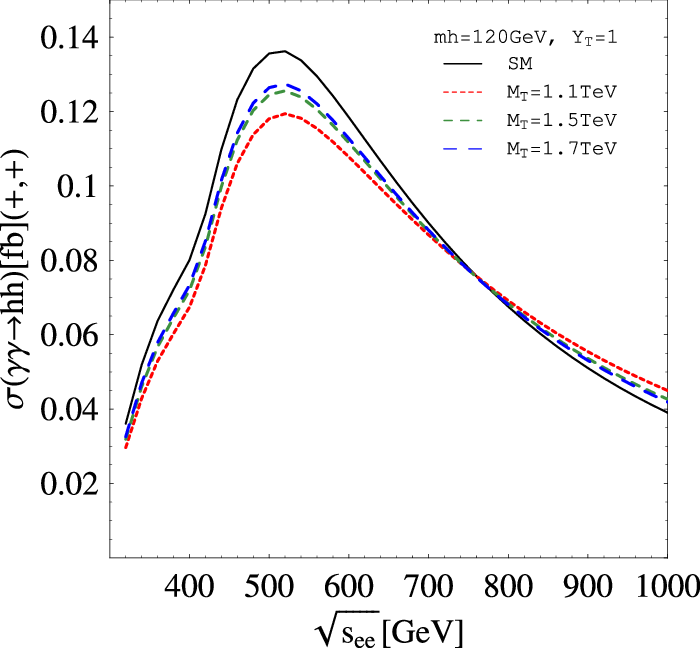}
\includegraphics[width=5.5cm]{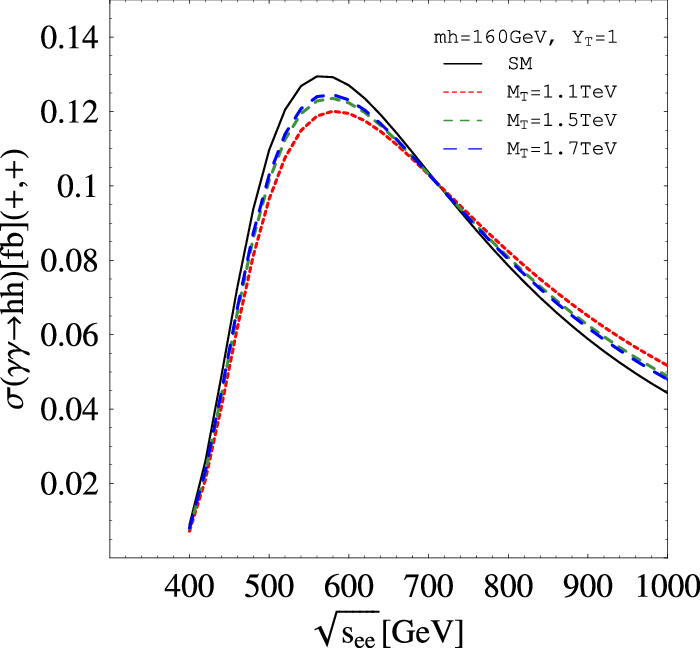}
\caption{The cross section of $\gamma(+)\gamma(+)\to hh$ process at the photon collider option
at the ILC as a function of the $e^-e^-$ collision energy for $m_h=120$ GeV (left) and $m_h=160$ GeV
(right) in the SM with the vectorlike top-quark.}
\label{FIG:GamGamhh-vf}
\end{figure}
In FIG.~\ref{FIG:GamGamhh-vf}, we show the cross section of $\gamma\gamma\to hh$ process in the model
with vectorlike top-quark. Similarly to the gluon fusion process $gg\to hh$, the new physics effects
are rather small.
%
\begin{figure}[tb]
\includegraphics[height=5.5cm]{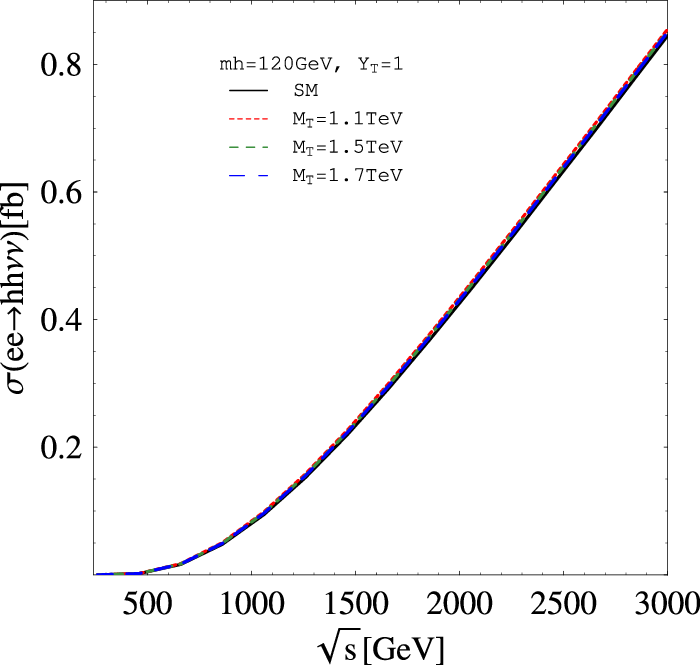}
\includegraphics[height=5.5cm]{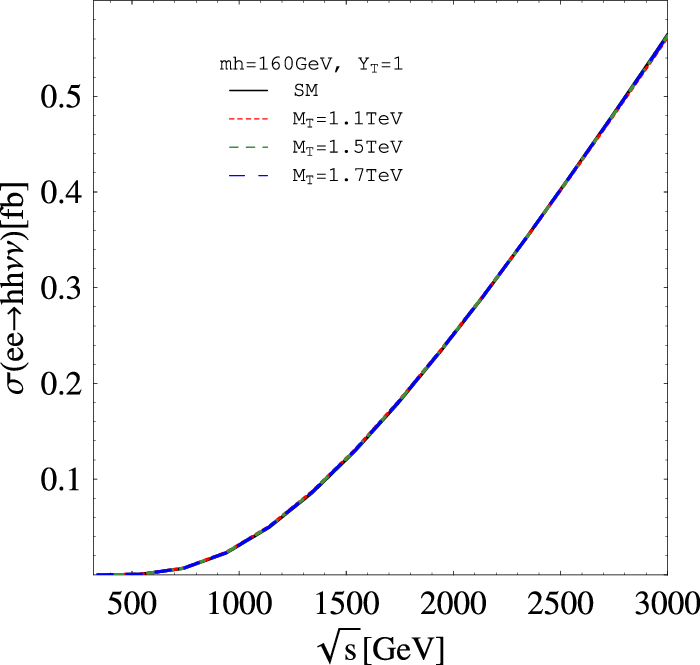}
\caption{The cross sections of $e^+e^-\to hh\nu\bar{\nu}$ process at the ILC as a function of collision
energy $\sqrt{s}$ for $m_h=120$ GeV (left) and $m_h=160$ GeV (right) in the SM with vectorlike top-quark.}
\label{FIG:eehhNNVec}
\end{figure}
In FIG.~\ref{FIG:eehhNNVec}, the cross section for $e^+e^-\to hh\nu\bar{\nu}$ is shown as a function of
$e^+e^-$ energy. Since the effect on the $hhh$ coupling constant are small in this model, the deviation
of cross section is very tiny.
We note that the deviations of cross sections for $gg\to hh$ and $\gamma\gamma\to hh$ are larger than
those for $e^+e^-\to hhZ$ and $e^+e^-\to hh\nu\bar\nu$ due to the flavor changing Yukawa interaction
between $t$ and $T$ from the one-loop box diagrams.

\section{Summary and discussions}
\begin{table}
\begin{tabular}{|c||c|c|c|c|}
\hline
$m_h$[GeV] & $\sigma_\text{SM}^{gg\to hh}$[fb]
& $\sigma_\text{SM}^{e^+e^-\to hhZ}$ [fb] & $\sigma_\text{SM}^{\gamma\gamma\to hh}$ [fb]
& $\sigma_\text{SM}^{e^+e^-\to hh\nu\bar\nu}$ [fb]\\
\hline\hline
$120$ & $21$  & $0.09$--$0.2$ & $0.14$            & $0.09$--$0.9$ \\ \hline
$160$ & $12$  & $0$--$0.05$   & $0.11$            & $0.02$--$0.6$ \\ \hline
$185$ & $8.0$ & $0$--$0.01$   & $0.07$            & $0.02$--$0.5$ \\ \hline
$210$ & $5.2$ & $0$           & $\thickapprox 0$ & $0.01$--$0.4$ \\ \hline
\end{tabular}
\caption{The total cross sections of Higgs boson pair production for $m_h=120, 160, 185$ and
$210$ GeV in the SM are listed. The proton-proton collision energy is taken as $\sqrt{s}=14$ TeV
for the gluon fusion mechanism. For the double-Higgs-strahlung $\sqrt{s_{ee}}$ is varied from $400$
GeV to $500$ GeV, while for the $W$ boson fusion $\sqrt{s_{ee}}$ is varied from $1$ TeV to $3$ TeV.
For photon-photon collisions energy is optimized to obtain the largest cross sections in a range
$\sqrt{s_{ee}}\gtrsim 500$ GeV.}
\label{TAB:summary1}
\end{table}

We have studied the double Higgs boson production processes $gg\to hh$, $e^+e^-\to hhZ$,
$\gamma\gamma\to hh$ and $e^+e^-\to hh\nu\bar\nu$ in various new physics models. These processes
include diagrams that contain the $hhh$ coupling constant, so that they can be used to obtain
information of the Higgs potential. SM cross sections for these processes are shown at the leading
order in Table~\ref{TAB:summary1} for several values of $m_h$. In order to see the impact of
a deviation in the $hhh$ coupling constant on these double Higgs boson production processes, we have
at first evaluated their cross sections in the SM but assuming the constant deviation in the $hhh$
coupling constant by the factor of $(1+\Delta\kappa)$. The results are summarized as follows:

\begin{itemize}
\item
The $gg\to hh$ process is the one-loop induced process, where contributions from the triangle
top-loop diagrams with the $hhh$ coupling constant and the box-type top-loop diagrams are destructive.
Therefore, a negative (positive) deviation in the $hhh$ coupling constant can make the cross section
larger (smaller). For example, when $\Delta\kappa=-1$, namely the case without the $hhh$ coupling,
the cross section can be approximately doubled (tripled) as compared to the SM value with
$\Delta\kappa=0$ for $m_h=120$  GeV (160  GeV).
\item
For the process of $e^+e^- \to hhZ$, on the other hand, the contribution of the tree-level diagram
with the $hhh$ coupling constant and that of the other tree-level diagrams are constructive, so that
the cross section is enhanced by the positive deviation in the $hhh$ coupling constant. Because of
an $s$-channel process, as seen in Table~\ref{TAB:summary1}, the cross section becomes rapidly smaller
for a larger mass of the Higgs boson, so that this process may only be useful for a light Higgs bosons
like $m_h\lesssim 140$ GeV for $\sqrt{s}\simeq 500$ GeV.
\item
The one-loop induced process $\gamma\gamma\to hh$ can play a complementary role to $gg\to hh$ and
$e^+e^-\to hhZ$. For $\sqrt{s_{ee}}\lesssim 500$ GeV, the sensitivity to the deviation in
the $hhh$ coupling constant is significant for both $m_h=120$ GeV and $160$ GeV. Direction of
interference of the diagram with the $hhh$ coupling constant and the other diagrams is however
opposite; i.e., destructive and constructive for $m_h=120$ GeV and $160$ GeV, respectively.
These characteristic behaviors of $\gamma\gamma\to hh$ can be complementary to $gg\to hh$ and
$e^+e^-\to hhZ$ in the measurement of the $hhh$ coupling constant. Furthermore, sensitivity to
the deviation in the $hhh$ coupling constant can be better by using this process even when the
collision energy of the linear collider is limited to be relatively low
($\sqrt{s_{ee}}\lesssim 500$ GeV).
\item
The $W$ boson fusion process $e^+e^-\to hh\nu\bar\nu$ can be useful for measuring the $hhh$ coupling
constant at an energy upgrade of the ILC or the CLIC with $\sqrt{s_{ee}}=1$--$3$ TeV, because the
production cross section is monotonically increasing with the collision energy due to the $t$-channel
colinear effect. The diagram with the $hhh$ coupling constant has the opposite sign with the
other diagrams.
\end{itemize}

We have evaluated cross sections for these processes in various new physics models such as the THDM,
the model with scalar leptoquarks, the model with chiral fourth generation quarks, and the model
with a vectorlike quark. In these models, apart from the deviated $hhh$ coupling constant, additional
one-loop diagrams can contribute to the cross sections especially in the loop induced processes
$gg\to hh$ and $\gamma\gamma\to hh$. Only one-loop diagrams of colored particles contribute to the
former process, while those of all the charged particles do to the latter one.
In Table~\ref{TAB:summary2}, the results for possible deviations in cross sections for these processes
are summarized with the deviation in the $hhh$ coupling constant for several typical values of $m_h$
and the collision energies in each model. We summarize the results for each model below in order.

\begin{table}
\begin{tabular}{|c||c|c|c|c|c|c|}
\hline Model
& $m_h$[GeV] & $\frac{\Gamma_{hhh}^\text{NP}-\Gamma_{hhh}^\text{SM}}{\Gamma_{hhh}^\text{SM}}$
& $\Delta r_\text{NP}^{gg\to hh}$ & $\Delta r_\text{NP}^{e^+e^-\to hhZ}$
& $\Delta r_\text{NP}^{\gamma\gamma\to hh}$ & $\Delta r_\text{NP}^{e^+e^-\to hh\nu\bar\nu}$\\
\hline\hline
THDM            & $120$ & $+120\%$ & $-50\%$
& $+(80$--$70)\%$   & $+50\%$         & $-(80$--$50)\%$   \\ \hline
THDM            & $160$ & $+70\%$ & $-50\%$   &
$+(60$--$50)\%$   & $+110\%$        & $-(80$--$50)\%$   \\ \hline
LQ($Q=1/3,4/3$) & $120$ & $+150\%$ & $-40\%$
& $+(110$--$100)\%$ & $+130\%,+100\%$   & $-(70$--$60)\%$   \\ \hline
LQ($Q=1/3,4/3$) & $185$ & $+60\%$ & $-30\%$   &
$+50\%$           & $+150\%,+150\%$ & $-(80$--$50)\%$   \\ \hline
Ch$4$             & $120$ & $-590\%$ & $+7800\%$
& $-(30$--$20)\%$   & $+3100\%$       & $+(260$--$110)\%$ \\ \hline
Ch$4$             & $210$ & $-140\%$ & $+2200\%$
& ------            & ------ & $+(970$--$210)\%$\\ \hline
Vec             & $120$ & $-4\%$ & $-10\%$   &
$-2\%$            & $-10\%$         & $+(5$--$1)\%$     \\ \hline
Vec             & $160$ & $-2\%$ & $-5\%$    &
$-1\%$            & $-10\%$         & $+(3$--$0)\%$\\ \hline
\end{tabular}
\caption{
Possible quantum corrections to the $hhh$ coupling constant,
$\frac{\Gamma_{hhh}^\text{NP}(4m_h^2,m_h^2,m_h^2)}{\Gamma_{hhh}^\text{SM}(4m_h^2,m_h^2,m_h^2)}-1$,
and deviations of cross sections
$\Delta r_\text{NP}^{}\equiv (\sigma_\text{NP}-\sigma_\text{SM})/\sigma_\text{SM}$ are listed.
The proton-proton collision energy is taken as $\sqrt{s}=14$ TeV for the gluon fusion mechanism.
For the double-Higgs-strahlung $\sqrt{s_{ee}}$ is varied from $400$ GeV to $500$ GeV, while for
the $W$ boson fusion $\sqrt{s_{ee}}$ is varied from $1$ TeV to $3$ TeV. For photon-photon collisions
$\sqrt{s_{ee}}$ is optimized to obtain the largest cross sections. Model parameters are chosen as
$m_\Phi=400$ GeV and $M^2=0$ for THDM, $m_\phi=400$ GeV and $M_\text{\lq}^2= 0$ for scalar leptoquark
models, $m_{t'}=400$ GeV and $m_{b'}=345$ GeV for the fourth generation model, and $Y_T=1, M_T=1100$
GeV for vectorlike top-quark model, respectively.}
\label{TAB:summary2}
\end{table}

In the THDM, one-loop corrections of additional scalar bosons to the $hhh$ coupling constant
can give $+100\%$ deviations in the SM-like limit where we set $\sin(\beta-\alpha)=1$ with
$M^2\simeq0$. This positive large quantum correction to the $hhh$ coupling constant is common
in the extended Higgs sectors with nondecoupling property where the mass of the scalar bosons
comes mainly from the VEV. The effect of the large deviation in the $hhh$ coupling constant
can be well described by the analysis with the constant shift of the $hhh$ coupling constant
by the factor of $(1+\Delta\kappa)$. A qualitative difference can be seen in the one-loop induced
$\gamma\gamma\to hh$ process, where one-loop diagrams of charged Higgs bosons can change to the
cross section.

In the scalar leptoquark models, the correction to the $hhh$ vertex is positive because of
additional bosonic loop contributions, as in the THDM. The magnitude can be larger than $+100\%$
via the nondecoupling effect of scalar leptoquarks in the loop when $M_\text{\lq}^2\simeq 0$.
A qualitative difference from the THDM case is that the scalar leptoquarks are colored, which
can contribute to the $gg\to hh$ through the one particle irreducible one-loop diagram. However,
it turns out that the top-quark one-loop contribution is much larger than the leptoquark-loop
contribution, so that the SM result with $\Delta\kappa$ correction is a good approximation.
Therefore, as expected in the analysis by using the $\Delta\kappa$, positive deviations in
the $hhh$ coupling constant make the cross sections smaller. It amounts to minus $40(30)\%$ for
$m_h=120(185)$ GeV assuming the other parameters as $m_\phi=400$ GeV and $M^2_\text{\lq}=0$.
The production rates for $e^+e^-\to hhZ$ can be enhanced by $+100(+50)\%$ for $m_h=120(185)$ GeV
due to the constructive interference in the contribution from the diagram with a positively
deviated $hhh$ coupling constant and the other diagrams. On the contrary, cross sections for
$e^+e^-\to hh\nu\bar\nu$ become smaller due to the destructive interference.
The production rates for $\gamma\gamma\to hh$ depend on electric charges of leptoquarks.
For the scalar leptoquark with $Q=4/3$, the cross section can be enhanced by the threshold effect
at $\sqrt{s_{\gamma\gamma}}\sim 2m_\phi$, while for those with $Q=1/3$ such effects are smeared.
For $\sqrt{s_{ee}}\gtrsim 500$ GeV, the cross section for the scalar leptoquarks with $Q=4/3$
can enhance more than several times $+100\%$.

In the model with chiral fourth generation quarks, the $hhh$ coupling constant can be changed
by more than a few times $-100\%$ due to the nondecoupling loop effect of additional heavy
chiral fermions. These huge corrections can be possible under the constraint from the data for
precision measurements and the direct search results. However, such a large fermionic loop
contribution can make Higgs potential unstable, so that a heavier Higgs boson is required than
the allowed value in the SM to recover the stability of vacuum. A light Higgs boson can be allowed
by extending the Higgs sector with additional scalar doublets.
The cross section for $gg\to hh$ is drastically enhanced from the SM prediction by a factor of
$10$--$100$, because new colored particles contribute to the additional one-loop diagrams at
leading order and the fourth generation fermions enhance both the one-loop induced vertex $ggh$
and the one-loop corrected $hhh$ coupling constant. For a reference point of $m_{t'}=m_{\ell}=400$
GeV and $m_{b'}=m_{\nu'}=345$ GeV, the cross sections become $7800$ $(2200)\%$ for $m_h=120$ $(210)$
GeV. Consequently, the process $gg\to hh$ with the decay mode of $h\to WW/ZZ$ can be promising
to measure the $hhh$ coupling constant for $m_h\gtrsim 210$ GeV.
For $e^+e^-\to hhZ$ process, the effect of fourth generation fermions only appear in the one-loop
corrected $hhh$ coupling constant. Since this correction is negative because of the fermionic loop
contribution, the cross section is suppressed. Numerically, it is reduced by $30$--$20\%$ for the
above reference point with $m_h=120$ GeV and $400$ GeV $\lsim \sqrt{s_{ee}} \lsim 500$ GeV.
On the contrary, in $e^+e^-\to hh\nu\bar\nu$ process, the effect of fourth generation fermions only
appears in the $hhh$ coupling constant through large one-loop corrections, which makes the cross section
huge as compared to the SM value.
Cross sections for $\gamma\gamma\to hh$ at photon colliders can also be modified by extra fermion
loops similarly to those for $gg\to hh$. With an optimized value of $\sqrt{s_{ee}}  \lsim 500$ GeV,
the cross section can be enhanced by $3100\%$ for the reference point with $m_h=120$ GeV.
Since the cross sections are drastically enhanced for $m_h=120$ GeV and hadronic decay modes
can be measurable at the PLC, the process $\gamma\gamma\to hh$ would be a promising process to probe
the $hhh$ coupling constant.

In the vectorlike quark model, a nondecoupling limit $(M_T\simeq 0)$ cannot be taken due to the severe
experimental constraints. Therefore, there are no large one-loop effect from a vector top-quark on
the $hhh$ coupling constant. The cross sections for $gg\to hh$ and $\gamma\gamma\to hh$ can be deviated
slightly from SM prediction due to the flavor changing Yukawa interaction of vectorlike quarks.
However, deviations of Higgs boson pair production cross sections are rather small, so that huge
luminosity would be required for measuring the deviation of the $hhh$ coupling constant.
%

Measuring four kinds of double Higgs boson production processes at different future collider
experiments is useful to discriminate whether new physics particles in the loop are fermions
or bosons and also whether they are colored or not.
Higgs boson pair production processes $e^+e^-\to hhZ$ and $e^+e^-\to hh\nu\bar\nu$ at lepton
colliders are tree-level processes, which enable us to extract information of the $hhh$ coupling
constant. On the other hand, measurements of effective vertices $gghh$ and $\gamma\gamma hh$ in
loop induced processes $gg\to hh$ at the LHC and $\gamma\gamma\to hh$ at the PLC can provide
information of colored and electrically charged particles in loop diagrams. Effective vertices
$ggh$ and $\gamma\gamma h$ can be determined in the single Higgs boson production processes
$gg\to h$ and $\gamma\gamma\to h$ as well. Combining these measurements, we would be able to
disentangle new physics effects in the $hhh$ coupling and the effective vertices.

We have considered the Higgs boson pair production processes, $gg\to hh$, $e^+e^-\to hhZ$,
$e^+e^-\to hh\nu\bar\nu$ and $\gamma\gamma\to hh$ as a probe of the $hhh$ coupling constant.
The measurement of the $hhh$ coupling constant is particularly important to understand
the mechanism of the electroweak symmetry breaking. The $hhh$ coupling constant can receive
quite large quantum corrections from new physics particles as a nondecoupling effect, which
can be an order of more than $100$\%. Deviations of the $hhh$ coupling constant can give different
effects on these processes which can largely modify production cross sections. Additional
particles in new physics model can also significantly affect the $gg\to hh$ and the
$\gamma\gamma\to hh$ processes according to their color and electric charges. We have found
that these four Higgs boson pair production processes at different colliders can play
complementary roles in exploring new physics through the Higgs sector.

\vspace{1cm} \noindent
{\bf Acknowledgments}~~~\\[2mm]
We would like to thank Kaoru Hagiwara for useful comments.
S.~K. acknowledges support from Grant-in-Aid for Science Research, Japan Society
for the Promotion of Science (JSPS), No.~$18034004$.
Y.~O.  acknowledges support from Grant-in-Aid for Science Research, MEXT-Japan,
No.~$16081211$, and JSPS, No.~$20244037$.
S.~K. and Y.~O. acknowledge support from Grant-in-Aid for Scientific Research
No.~$22244031$ of JSPS.
Numerical calculations were partly carried out on Altix$3700$ at Yukawa Institute for 
Theoretical Physics (YITP) in Kyoto University.

\appendix
\section{The one-loop corrections to the triple Higgs boson coupling}
The relatively large one-loop correction to the $hhh$ coupling constant in the SM has been
calculated as~\cite{Ref:KOSY},
\begin{align}
\frac{\Gamma^\text{SM}_{hhh}(p_1^2,p_2^2,q^2)}{\lambda_{hhh}^\text{SM}}
=1-\frac{N_c}{16\pi^2}\Biggl\{
&+\sum_{f=t, b}\frac{m_f^2}{m_h^2v^2}(-2m_h^2+8m_f^2)B_0(m_h^2;m_f,m_f)\nonumber\\
&\hspace{-5ex}
-\sum_{(f_1,f_2)=(t, b)}\frac4{v^2}\left[B_{22}(0;m_{f_1},m_{f_2})-\frac14(m_{f_1}^2+m_{f_2}^2)\right]\nonumber\\
&\hspace{-5ex}
+\sum_{f=t, b}\frac{3m_f^2}{2v^2}\left[2B_0(m_h^2;m_f,m_f)-(-2m_h^2+8m_f^2)B'_0(m_h^2;m_f,m_f)\right]\nonumber\\
&\hspace{-5ex}
-\sum_{f=t, b}\frac{8m_f^4}{3m_h^2v^2}\Bigl[B_0(p_1^2;m_f,m_f)+B_0(p_2^2;m_f,m_f)+B_0(q^2;m_f,m_f)\nonumber\\
&\hspace{-5ex}\quad-\frac12(p_1^2+p_2^2+q^2-8m_f^2)C_0(p_1^2,p_2^2,q^2;m_f,m_f,m_f)\Bigr]
\Biggr\},
\end{align}
where $B$ and $C$ are the loop functions, which are defined in Ref.~\cite{Ref:PV}.
For the chiral fourth generation model, all the extra fermions further contribute to
the $hhh$ coupling constant. Then the sum should be replaced by all heavy fermions
in the preceding formula.

In the leptoquark model, the quantum effect is given by
\begin{align}
\frac{\Gamma^\text{\lq}_{hhh}(p_1^2,p_2^2,q^2)}{\lambda^\text{SM}_{hhh}}
=&\frac{\Gamma^\text{SM}_{hhh}(p_1^2,p_2^2,q^2)}{\lambda^\text{SM}_{hhh}}
+\frac{N_c}{16\pi^2}\Biggl\{
+\frac1{m_h^2}\lambda_{h\phi\phi^*}^2\,B_0(m_h^2;m_{\phi}^{},m_{\phi}^{})
-\frac32\lambda_{h\phi\phi^*}^2\,B'_0(m_h^2;m_{\phi}^{},m_{\phi}^{})\nonumber\\
&-\frac{v}{3m_h^2}\Bigl[
-2\lambda_{h\phi\phi^*}^3C_0(p_1^2,p_2^2,q^2;m_{\phi}^{},m_{\phi}^{},m_{\phi}^{})\nonumber\\
&\quad+2\lambda_{h\phi\phi^*}^{}\frac{\lambda_{hh\phi\phi^*}^{}}{v}
\left[B_0(q^2;m_{\phi}^{},m_{\phi}^{})+B_0(p_1^2;m_{\phi}^{},m_{\phi}^{})
+B_0(p_2^2;m_{\phi}^{},m_{\phi}^{})\right]\Bigr]\Biggr\}.
\end{align}

In the THDM with the SM-like limit $\sin(\beta-\alpha)=1$, the one-loop corrections to
the $hhh$ coupling constant are calculated as~\cite{Ref:KOSY},
\begin{align}
&\frac{\Gamma^\text{THDM}_{hhh}(p_1^2,p_2^2,q^2)}{\lambda^\text{SM}_{hhh}}
=\frac{\Gamma^\text{SM}_{hhh}(p_1^2,p_2^2,q^2)}{\lambda^\text{SM}_{hhh}}
+\frac1{16\pi^2}\Biggl\{\nonumber\\
&+\frac1{m_h^2}\left[\lambda_{hH^+H^-}^2\,B_0(m_h^2;m_{H^\pm}^{},m_{H^\pm}^{})
+2\lambda_{hHH}^2\,B_0(m_h^2;m_H^{},m_H^{})+2\lambda_{hAA}^2\,B_0(m_h^2;m_A^{},m_A^{})\right]\nonumber\\
&-\frac1{2v^2}\left[B_5(0;m_H^{},m_{H^\pm}^{})+B_5(0;m_A^{},m_{H^\pm}^{})\right]\nonumber\\
&-\frac32\left[\lambda_{hH^+H^-}^2\,B'_0(m_h^2;m_{H^\pm}^{},m_{H^\pm}^{})
+2\lambda_{hHH}^2\,B'_0(m_h^2;m_H^{},m_H^{})+2\lambda_{hAA}^2\,B'_0(m_h^2;m_A^{},m_A^{})\right]\nonumber\\
&-\frac{v}{3m_h^2}\Biggl[
-2\lambda_{hH^+H^-}^3C_0(p_1^2,p_2^2,q^2;m_{H^\pm}^{},m_{H^\pm}^{},m_{H^\pm}^{})\nonumber\\
&\quad-8\lambda_{hHH}^3C_0(p_1^2,p_2^2,q^2;m_H^{},m_H^{},m_H^{})
-8\lambda_{hAA}^3C_0(p_1^2,p_2^2,q^2;m_A^{},m_A^{},m_A^{})\nonumber\\
&\quad+2\lambda_{hH^+H^-}^{}\frac{\lambda_{hhH^+H^-}^{}}{v}
\left[B_0(q^2;m_{H^\pm}^{},m_{H^\pm}^{})+B_0(p_1^2;m_{H^\pm}^{},m_{H^\pm}^{})
+B_0(p_2^2;m_{H^\pm}^{},m_{H^\pm}^{})\right]\nonumber\\
&\quad+4\lambda_{hHH}^{}\frac{\lambda_{hhHH}^{}}{v}
\left[B_0(q^2;m_H^{},m_H^{})+B_0(p_1^2;m_H^{},m_H^{})
+B_0(p_2^2;m_H^{},m_H^{})\right]\nonumber\\
&\quad+4\lambda_{hAA}^{}\frac{\lambda_{hhAA}^{}}{v}
\left[B_0(q^2;m_A^{},m_A^{})+B_0(p_1^2;m_A^{},m_A^{})
+B_0(p_2^2;m_A^{},m_A^{})\right]\Biggr]
\Biggr\},
\end{align}
where
\begin{align}
\lambda_{hH^+H^-}^{}&=2\lambda_{hhH^+H^-}^{}=-\frac{m_h^2}{v}
-\frac{2m_{H^\pm}^2}{v}\left(1-\frac{M^2}{m_{H^\pm}^2}\right),\\
\lambda_{hHH}^{}&=2\lambda_{hhHH}^{}=\frac12\left[-\frac{m_h^2}{v}
-\frac{2m_H^2}{v}\left(1-\frac{M^2}{m_H^2}\right)\right],\\
\lambda_{hAA}^{}&=2\lambda_{hhAA}^{}=\frac12\left[-\frac{m_h^2}{v}
-\frac{2m_A^2}{v}\left(1-\frac{M^2}{m_A^2}\right)\right].
\end{align}

In the vectorlike top-quark model, the one-loop corrected $hhh$ coupling constant is given by
\begin{align}
&\frac{\Gamma^\text{Vec}_{hhh}(p_1^2,p_2^2,q^2)}{\lambda^\text{SM}_{hhh}}
=1-\frac{N_c}{16\pi^2}\Biggl\{\nonumber\\
&+\frac1{m_h^2}
\Bigl\{-4\frac{m_f}{v}\left(\frac{y_t^\text{eff}}{\sqrt2}\right)A(m_t)
+\left(\frac{y_t^\text{eff}}{\sqrt2}\right)^2\bigl[4A(m_t)+(-2m_h^2+8m_t^2)B_0(m_h^2;m_t,m_t)\bigr]\nonumber\\
&\quad-4\frac{m_T^{}}{v}\left(\frac{y_T^\text{eff}}{\sqrt2}\right)A(m_T^{})
+\left(\frac{y_T^\text{eff}}{\sqrt2}\right)^2
\bigl[4A(m_T^{})+(-2m_h^2+8m_T^2)B_0(m_h^2;m_T^{},m_T^{})\bigr]\nonumber\\
&\quad+2\left[\left(\frac{\epsilon_t}{\sqrt2}\right)^2+\left(\frac{\epsilon_T^{}}{\sqrt2}\right)^2\right]
\left[A(m_t)+A(m_T^{})+(-m_h^2+m_t^2+m_T^2)B_0(m_h^2;m_t,m_T^{})\right]\nonumber\\
&\quad+8\left(\frac{\epsilon_t}{\sqrt2}\right)\left(\frac{\epsilon_T^{}}{\sqrt2}\right)m_tm_T^{}
B_0(m_h^2;m_t,m_T^{})\Bigr\}\nonumber\\
&-\frac1{v^2}\left[4c_{L}^2B_{22}(0;m_t,0)-c_{L}^2m_t^2
+4s_{L}^2B_{22}(0;m_T^{},0)-s_{L}^2m_T^2\right]\nonumber\\
&+\frac32\Bigl\{+\left(\frac{y_t^\text{eff}}{\sqrt2}\right)^2\bigl[2B_0(m_h^2;m_t,m_t)
-(-2m_h^2+8m_t^2)B'_0(m_h^2;m_t,m_t)\bigr]\nonumber\\
&\quad+\left(\frac{y_T^\text{eff}}{\sqrt2}\right)^2
\bigl[2B_0(m_h^2;m_T^{},m_T^{})-(-2m_h^2+8m_T^2)B'_0(m_h^2;m_T^{},m_T^{})\bigr]\nonumber\\
&\quad+2\left[\left(\frac{\epsilon_t}{\sqrt2}\right)^2+\left(\frac{\epsilon_T^{}}{\sqrt2}\right)^2\right]
\left[B_0(m_h^2;m_t,m_T^{})-(-m_h^2+m_t^2+m_T^2)B'_0(m_h^2;m_t,m_T^{})\right]\nonumber\\
&\quad-8\left(\frac{\epsilon_t}{\sqrt2}\right)\left(\frac{\epsilon_T^{}}{\sqrt2}\right)
m_tm_T^{}B'_0(m_h^2;m_t,m_T^{})\Bigr\}\nonumber\\
&-\frac{v}{3m_h^2}\Biggl\{8m_t\left(\frac{y_t^\text{eff}}{\sqrt2}\right)^3
\Bigl[+B_0(p_1^2;m_t,m_t)+B_0(p_2^2;m_t,m_t)+B_0(q^2;m_t,m_t)\nonumber\\
&\qquad-\frac12(p_1^2+p_2^2+q^2-8m_t^2)C_0(p_1^2,p_2^2,q^2;m_t,m_t,m_t)\Bigr]\nonumber\\
&\quad+4m_t\left(\frac{y_t^\text{eff}}{\sqrt2}\right)
\left[\left(\frac{\epsilon_t}{\sqrt2}\right)^2+\left(\frac{\epsilon_T^{}}{\sqrt2}\right)^2\right]\nonumber\\
&\qquad\times\Bigl[+B_0(p_1^2;m_t,m_t)+B_0(p_2^2;m_t,m_t)+B_0(q^2;m_t,m_t)\nonumber\\
&\hspace{10ex}+B_0(p_1^2;m_t,m_T^{})+B_0(p_2^2;m_t,m_T^{})+B_0(q^2;m_t,m_T^{})\nonumber\\
&\hspace{10ex}-\frac12(p_1^2+p_2^2-2m_t^2-2m_T^2)C_0(p_1^2,p_2^2,q^2;m_t,m_T^{},m_t)\nonumber\\
&\hspace{10ex}-\frac12(p_1^2+q^2-2m_t^2-2m_T^2)C_0(p_1^2,p_2^2,q^2;m_T^{},m_t,m_t)\nonumber\\
&\hspace{10ex}-\frac12(p_2^2+q^2-2m_t^2-2m_T^2)C_0(p_1^2,p_2^2,q^2;m_t,m_t,m_T^{})\Bigr]\nonumber\\
&\quad+8m_T^{}\left(\frac{y_t^\text{eff}}{\sqrt2}\right)
\left(\frac{\epsilon_t}{\sqrt2}\right)\left(\frac{\epsilon_T^{}}{\sqrt2}\right)\nonumber\\
&\qquad\times\Bigl[+B_0(p_1^2;m_t,m_T^{})+B_0(p_2^2;m_t,m_T^{})+B_0(q^2;m_t,m_T^{})\nonumber\\
&\hspace{10ex}-\frac12(q^2-4m_t^2)C_0(p_1^2,p_2^2,q^2;m_t,m_T^{},m_t)\nonumber\\
&\hspace{10ex}-\frac12(p_2^2-4m_t^2)C_0(p_1^2,p_2^2,q^2;m_T^{},m_t,m_t)\nonumber\\
&\hspace{10ex}-\frac12(p_1^2-4m_t^2)C_0(p_1^2,p_2^2,q^2;m_t,m_t,m_T^{})\Bigr]\nonumber\\
&+\left(y_t^\text{eff}\to y_T^\text{eff},m_t\leftrightarrow m_T^{}\right)\Biggr\}
\Biggr\},
\end{align}
where
\begin{align}
y_t^\text{eff}=&c_L^{}(c_R^{}y_t-s_R^{}Y_T),\\
\epsilon_t=&s_L^{}(c_R^{}y_t-s_R^{}Y_T),\\
y_T^\text{eff}=&s_L^{}(s_R^{}y_t+c_R^{}Y_T),\\
\epsilon_T^{}=&c_L^{}(s_R^{}y_t+c_R^{}Y_T).
\end{align}

\section{The loop integrals for $gg\to hh$ and $\gamma\gamma\to hh$}
The SM contributions to the loop functions in $gg\to hh$ and $\gamma\gamma\to hh$
amplitudes can be found in Refs.~\cite{Ref:hhh-sensitivity,Ref:hhhPLC}. It can
be generalized straightforwardly for the chiral fourth generation model.

In leptoquark models, in addition to the SM fermions there are contributions from
the colored scalar particle to $gg\to hh$ amplitude as,
\begin{align}
F_\triangle^\phi
=& \frac{4\lambda_{hh\phi\phi^*}}{v}\left(1-\frac{{\hat s}-m_h^2}{\lambda_{hhh}v}\right)
\left(1+2m_\phi^2C_0^{(1,2)}\right),\label{A1}\\
F_\Box^\phi
=&+4\lambda_{h\phi\phi^*}^2\Biggl\{
m_\phi^2\left(D_0^{(1,2,3)}+D_0^{(2,1,3)}+D_0^{(1,3,2)}\right)
\nonumber \\
&\qquad+\frac{{\hat t}{\hat u}-m_h^4}{2{\hat s}}D_0^{(1,3,2)}
-\frac{{\hat u}-m_h^2}{\hat s}C_0^{(2,3)}
-\frac{{\hat t}-m_h^2}{\hat s}C_0^{(1,3)}\Biggr\}, \label{A2}\\
G_\Box^\phi
=&+4\lambda_{h\phi\phi^*}^2
\Biggl\{-C_0^{(3,4)}
+m_\phi^2\left(D_0^{(1,2,3)}+D_0^{(2,1,3)}+D_0^{(1,3,2)}\right)\nonumber \\
&\quad+\frac1{2\left({\hat t}{\hat u}-m_h^4\right)}\Biggl[
-2{\hat u}({\hat u}-m_h^2)C_0^{(2,3)}-2{\hat t}({\hat t}-m_h^2)C_0^{(1,3)}
\nonumber\\
&\qquad+{\hat s}{\hat u}^2D_0^{(1,2,3)}+{\hat s}{\hat t}^2D_0^{(2,1,3)}
+{\hat s}({\hat s}-2m_h^2)C_0^{(1,2)}+{\hat s}({\hat s}-4m_h^2)C_0^{3,4}\Biggr]
\Biggr\},\label{A3}
\end{align}
where
\begin{align}
\lambda_{h\phi\phi^*}^{}
&=2\lambda_{hh\phi\phi^*}^{}
=-\frac{2m_\phi^2}{v}\left(1-\frac{M^2_\text{\lq}}{m_\phi^2}\right),
\end{align}
and the loop functions are abbreviated as
\begin{align}
C^{(1,2)}_{A[i,j,k]}&=C_A^{}(0,0,{\hat s};m_i,m_j,m_k),\label{Eq:C12}\\
C^{(3,4)}_{A[i,j,k]}&=C_A^{}({\hat s},m_h^2,m_h^2;m_i,m_j,m_k),\label{Eq:C34}\\
C^{(1,3)}_{A[i,j,k]}&=C_A^{}({\hat t},0,m_h^2;m_i,m_j,m_k),\label{Eq:C13}\\
C^{(2,3)}_{A[i,j,k]}&=C_A^{}(m_h^2,0,{\hat u};m_i,m_j,m_k),\label{Eq:C23}\\
D^{(1,2,3)}_{A[i,j,k,l]}&=D_A^{}(0,0,m_h^2,m_h^2,{\hat s},{\hat u};m_i,m_j,m_k,m_l)\label{Eq:D123},\\
D^{(2,1,3)}_{A[i,j,k,l]}&=D_A^{}(0,0,m_h^2,m_h^2,{\hat s},{\hat t};m_i,m_j,m_k,m_l)\label{Eq:D212},\\
D^{(1,3,2)}_{A[i,j,k,l]}&=D_A^{}(0,m_h^2,0,m_h^2,{\hat t},{\hat u};m_i,m_j,m_k,m_l)\label{Eq:D132}.
\end{align}
The index $A$ denotes classes of loop functions, and we omitted obvious mass indices of
intermediate particles in the Eqs.\eqref{A1}--\eqref{A3}. The leptoquarks also contribute
to $\gamma\gamma\to hh$ process. These additional contributions to the amplitude are
basically the same as those loop integrals for $gg\to hh$ amplitude except for electric
charges and a color factor;
$H^{++}_\phi=N_c^\phi Q_\phi^2F^\phi$ and $H^{+-}_\phi=N_c^\phi Q_\phi^2G^\phi$.

The SM contribution can be modified in the vectorlike quark model because of $t$ and $T$ mixing.
The vectorlike top-quark $T$ also gives additional contributions to $gg\to hh$ and
$\gamma\gamma\to hh$ amplitudes as
\begin{align}
&F_\triangle^\text{Vec}
=2\left(\frac{y_t^\text{eff}}{\sqrt2}\right)\left(\frac{m_t}{v}\right)
\left[-4B_0^{(1+2)}+16C_{24}^{(1,2)}-2{\hat s}C_0^{(1,2)}\right]_t
+\left(y_t^\text{eff}\to y_T^\text{eff},m_t\leftrightarrow m_T^{}\right),
\nonumber\\
&F_\Box^\text{Vec}+G_\Box^\text{Vec}
=-2\left(\frac{y_t^\text{eff}}{\sqrt2}\right)^2\Biggl\{
-4B_0^{(1+2)}+16C_{24}^{(1,2)}-2{\hat s}C_0^{(1,2)}\nonumber\\
&\hspace{30ex}
+{\hat s}\left[\left({\hat u}-4m_t^2\right)D_0^{(1,2,3)}
+\left({\hat t}-4m_t^2\right)D_0^{(2,1,3)}\right]\nonumber\\
&\hspace{30ex}
-\left[\left({\hat t}-m_h^2\right)\left({\hat u}-m_h^2\right)
+{\hat s}\left(4m_t^2-m_h^2\right)\right]D_0^{(1,3,2)}\nonumber\\
&\hspace{20ex}
+4\left({\hat s}-2m_h^2+8m_t^2\right)
\left[D_{27}^{(1,2,3)}+D_{27}^{(2,1,3)}+D_{27}^{(1,3,2)}
-\frac12C_0^{(3,4)}\right]\Biggr\}_t\nonumber\\
&\hspace{15ex}
-\left[\left(\frac{\epsilon_t}{\sqrt2}\right)^2
+\left(\frac{\epsilon_T^{}}{\sqrt2}\right)^2\right]\Biggl\{
-4B_{0[t,t]}^{(1+2)}+16C_{24[t,t,t]}^{(1,2)}-2{\hat s}C_{0[t,t,t]}^{(1,2)}\nonumber\\
&\hspace{20ex}
+{\hat s}\left[\left({\hat u}-m_t^2-m_T^2\right)D_{0[t,t,t,T]}^{(1,2,3)}
+\left({\hat t}-m_t^2-m_T^2\right)D_{0[t,t,t,T]}^{(2,1,3)}\right]\nonumber\\
&\hspace{20ex}
-\left[\left({\hat t}-m_h^2\right)\left({\hat u}-m_h^2\right)
+{\hat s}\left(m_t^2+m_T^2-m_h^2\right)\right]D_{0[t,T,t,T]}^{(1,3,2)}
\nonumber\\
&\hspace{10ex}+4\left({\hat s}-2m_h^2+2m_t^2+2m_T^2\right)
\left(D_{27[t,t,t,T]}^{(1,2,3)}+D_{27[t,t,t,T]}^{(2,1,3)}+D_{27[t,T,t,T]}^{(1,3,2)}
-\frac12C_{0[t,t,T]}^{(3,4)}\right)
\Biggr\}\nonumber\\
&\hspace{15ex}
+4\left(\frac{\epsilon_t}{\sqrt2}\right)\left(\frac{\epsilon_T^{}}{\sqrt2}\right)m_tm_T^{}
\Biggl\{{\hat s}\left(D_{0[t,t,t,T]}^{(1,2,3)}+D_{0[t,t,t,T]}^{(2,1,3)}+D_{0[t,T,t,T]}^{(1,3,2)}\right)
\nonumber\\
&\hspace{20ex}
-4\left(D_{27[t,t,t,T]}^{(1,2,3)}+D_{27[t,t,t,T]}^{(2,1,3)}+D_{27[t,T,t,T]}^{(1,3,2)}
-C_{0[t,t,T]}^{(3,4)}\right)
\Biggr\}\nonumber\\
&\hspace{20ex}
+\left(y_t^\text{eff}\to y_T^\text{eff},m_t\leftrightarrow m_T^{}\right),\\
&G_\Box^\text{Vec}=-\frac{{\hat t}{\hat u}-m_h^4}{2{\hat s}}\Biggl\{
-8\left(\frac{y_t^\text{eff}}{\sqrt2}\right)^2\Biggl[
\left({\hat s}-2m_h^2+8m_t^2\right)
\left[D_{23}^{(1,2,3)}+D_{23}^{(2,1,3)}+D_{12}^{(1,3,2)}+D_{22}^{(1,3,2)}\right]
\nonumber\\
&\hspace{10ex}+{\hat s}\left(D_{13}^{(1,2,3)}+D_{13}^{(2,1,3)}\right)
-(C_0^{(1,3)}+C_0^{(2,3)}+C_{11}^{(1,3)}+C_{11}^{(2,3)}
+\overline{C}_{12}^{(1,3)}+\overline{C}_{12}^{(2,3)})\Biggr]_t\nonumber\\
&\hspace{13ex}
-4\left[\left(\frac{\epsilon_t}{\sqrt2}\right)^2
+\left(\frac{\epsilon_T^{}}{\sqrt2}\right)^2\right]\Biggl[
+{\hat s}\left(D_{13[t,t,t,T]}^{(1,2,3)}+D_{13[t,t,t,T]}^{(2,1,3)}\right)\nonumber\\
&\hspace{10ex}+\left({\hat s}-2m_h^2+2m_t^2+2m_T^2\right)
\left(D_{23[t,t,t,T]}^{(1,2,3)}+D_{23[t,t,t,T]}^{(2,1,3)}
+D_{12[t,T,t,T]}^{(1,3,2)}+D_{22[t,T,t,T]}^{(1,3,2)}\right)
\nonumber\\
&\hspace{10ex}
-\left(C_{0[t,t,T]}^{(1,3)}+C_{0[t,t,T]}^{(2,3)}+C_{11[t,t,T]}^{(1,3)}+C_{11[t,t,T]}^{(2,3)}
+\overline{C}_{12[t,t,T]}^{(1,3)}+\overline{C}_{12[t,t,T]}^{(2,3)}\right)\Biggr]\nonumber\\
&\hspace{13ex}
-16\left(\frac{\epsilon_t}{\sqrt2}\right)\left(\frac{\epsilon_T^{}}{\sqrt2}\right)m_tm_T^{}
\left(D_{23[t,t,t,T]}^{(1,2,3)}+D_{23[t,t,t,T]}^{(2,1,3)}
+D_{12[t,T,t,T]}^{(1,3,2)}+D_{22[t,T,t,T]}^{(1,3,2)}\right)\nonumber\\
&\hspace{20ex}
\Biggr\}+\left(y_t^\text{eff}\to y_T^\text{eff},m_t\leftrightarrow m_T^{}\right),
\end{align}
where
\begin{align}
B_{A[i,j]}^{(1+2)}&=B_A({\hat s};m_i,m_j),\\
\overline{C}_{A[i,j,k]}^{(1,3)}&=C_A(0,{\hat t},m_h^2;m_i,m_j,m_k),\\
\overline{C}_{A[i,j,k]}^{(2,3)}&=C_A(0,m_h^2,{\hat u};m_i,m_j,m_k).
\end{align}
The loop functions for photon collision are expressed as $H^{++}_\text{Vec}=N_c^t Q_t^2F^\text{Vec}$
and $H^{+-}_\text{Vec}=N_c^t Q_t^2G^\text{Vec}$.

\end{document}